\documentclass[pra,letterpaper,showpacs,superscriptaddress,floatfix]{revtex4}

\usepackage{graphicx,psfrag,amsmath,amssymb,amsfonts,latexsym,color,dcolumn,bm}

\begin{document}

\title{Results from electrostatic calibrations for measuring \\ 
the Casimir force in the cylinder-plane geometry}

\author{Q. Wei}
\affiliation{Department of Physics and Astronomy,Dartmouth College,6127 Wilder Laboratory,Hanover,NH 03755,USA}

\author{D. A. R. Dalvit} 
\affiliation{Theoretical Division, MS B213, Los Alamos National Laboratory, Los Alamos, NM 87545, USA}

\author{F. C. Lombardo}
\affiliation{Departamento de Fisica J.J. Giambiagi, Facultad de Ciencias Exactas y 
Naturales, Universidad de Buenos Aires, Ciudad Universitaria, Pabell\'on 1, 1428 Buenos Aires, Argentina}

\author{F. D. Mazzitelli}
\affiliation{Departamento de Fisica J.J. Giambiagi, Facultad de Ciencias Exactas y 
Naturales, Universidad de Buenos Aires, Ciudad Universitaria, Pabell\'on 1, 1428 Buenos Aires, Argentina}

\author{R. Onofrio}
\affiliation{Dipartimento di Fisica ``Galileo Galilei'',Universit\`a  di Padova,Via Marzolo 8,Padova 35131,Italy}

\affiliation{Department of Physics and Astronomy,Dartmouth College,6127 Wilder Laboratory,Hanover,NH 03755,USA}

\date{\today}

\begin{abstract}
We report on measurements performed on an apparatus aimed to study 
the Casimir force in the cylinder-plane configuration. 
The electrostatic calibrations evidence anomalous behaviors in the 
dependence of the electrostatic force and the minimizing potential upon distance. 
We discuss analogies and differences of these anomalies with 
respect to those already observed in the sphere-plane configuration. 
At the smallest explored distances we observe frequency shifts of non-Coulombian 
nature preventing the measurement of the Casimir force in the same range. 
We also report on measurements performed in the parallel plane configuration, 
showing that the dependence on distance of the minimizing potential, if 
present at all, is milder than in the sphere-plane or cylinder-plane geometries. 
General considerations on the interplay between the distance-dependent minimizing 
potential and the precision of Casimir force measurements in the range relevant 
to detect the thermal corrections for all geometries are finally reported.
\end{abstract}


\maketitle

\section{Introduction}

Casimir forces \cite{Casimir} have been investigated since their inception 
as a macroscopic test of the irreducible fluctuations associated to quantum fields. 
They are geometrical in character, as they originate from the possibility to 
confine and shape the energy density of quantum fluctuations using proper boundary conditions. 
So far, the experimental attention has been mainly focused on the original parallel 
plate configuration \cite{Sparnaay,Bressi}, and the sphere-plane geometry 
\cite{vanBlockland,Lamoreaux,Mohideen,Chan,Decca,Decca1}, apart from the only 
experiment performed in a crossed-cylinder configuration \cite{Ederth}. 
Further motivations to pursue precision studies of the Casimir forces 
are related to the possibility to discover new forces of strength
similar or larger than the Newtonian gravitational coupling but 
with short-range or different distance scaling and expected to act, 
according to various models, in the submillimeter range \cite{Giudice}. 
This can be considered part of a broader program aimed at testing deviations 
from Newtonian gravitation in the nonrelativistic limit \cite{Fujii,Fischbach}. 
In the micrometer range the dominant source of background to non-Newtonian 
gravitational forces is provided by Casimir  forces \cite{Kuzmin,Mostesoko} 
(see also \cite{Reynaudrev,Onofrio} for reviews) and to discover such forces, or at 
least to provide reliable limits on their existence, one must control at 
the highest level of accuracy all systematic sources of deviation from 
the idealized case analyzed by Casimir in his original paper. 
Major sources of systematic errors that are still considered under partial 
control  are the use of the proximity force approximation (PFA) 
\cite{Derjaguin,Blocki} for curved surfaces, the presence of electric 
forces not reflected in the purely Coulombian contribution, such as 
patch effects \cite{Burnham,Stipe,Speake,Kimpatch2,Podgornik}, and 
the combined effect of  finite conductivity and finite temperature 
\cite{Mehra,Brown,Bostrom}.

The use of the PFA  has been discussed at length in the literature,
with several alternative methods developed to overcome its limitations, 
and exact solutions have been found in particular curved geometries. 
It is generally assumed that the PFA for the Casimir force differs 
from the exact result by an amount smaller than 0.1 $\%$, an assumption 
compatible with the results of a dedicated experiment \cite{DeccaPFAexp}. 
The control on the PFA used to assess limits to Yukawian non-Newtonian gravity has 
not been addressed until very recently, although it has been used for many years 
\cite{Fischbach1,Decca2003,Decca1,Klim,Decca2007,Decca2007bis,MostepanenkoJPA}. 
It has been argued that the usual form of PFA cannot be extended unambiguously 
to volumetric forces \cite{ReplyPRARC}, and thereafter an alternative form of the 
PFA has been discussed \cite{DeccaPFA}, which, however, has been shown in 
\cite{DiegoRoberto} to coincide with the exact formula for geometries in 
which one of the two bodies has translational invariance. 

The presence of electric forces not incorporated in the Coulombian contribution 
has been discussed extensively in the literature, in particular in \cite{Burnham}.
Anomalies in the electrostatic calibrations of the sphere-plane configuration 
have been evidenced for large radii of curvature of the sphere and small 
distances from the planar surface \cite{PRARC,JPCS}. 
This has triggered discussions about the nature and the universality of the observed 
anomalies \cite{DeccaComment,ReplyPRARC}, and the situation is still far from being clarified. 
The anomalous exponent optimizing the fit of the electrostatic calibrations in
\cite{PRARC,JPCS} has {\sl not} been found in another experiment using spheres 
of much smaller diameter located at similar distances from the planar surface 
\cite{Iannuzzi}, while the dependence of the minimizing potential on the 
sphere-plane separation reported in \cite{PRARC} has been confirmed in 
\cite{Iannuzzi}, and for crystalline Ge in \cite{Kimpatch1} 
(see also \cite{JPCS} for a discussion of some unpublished data from former experiments).
The spatial and temporal variabilities of the minimizing potential has been 
evidenced in a centimeter-size torsional balance \cite{Pollack}, which confirms the necessity for a detailed 
knowledge of the surfaces and their preparation \cite{Schroder,Camp1,Camp2,Camp3}.

The finite temperature contribution added to the quantum fluctuations 
has originated a lengthy debate about the interplay of the thermal 
contribution with the finite conductivity properties of the surfaces 
(see for instance \cite{t1,t2,t3,t4,t5,t6,t7,t8,t9,t10,t11,t12} for 
the initial steps of the debate). On the experimental side, attempts 
to evidence the thermal contribution discriminating various models 
have been reported for the sphere-plane geometry \cite{Decca1}, 
while proposals using torsional balances in the parallel plane 
configuration \cite{Buttler,LambrechtCQG} are under development. 
A dedicated experiment in the parallel plane configuration using 
microresonators \cite{Antonini1} and a low-frequency heterodyne 
technique \cite{Bressicqg} has been limited so far from patch 
charges \cite{Antonini2}. 

\begin{figure}[b]
\includegraphics[width=0.45\columnwidth]{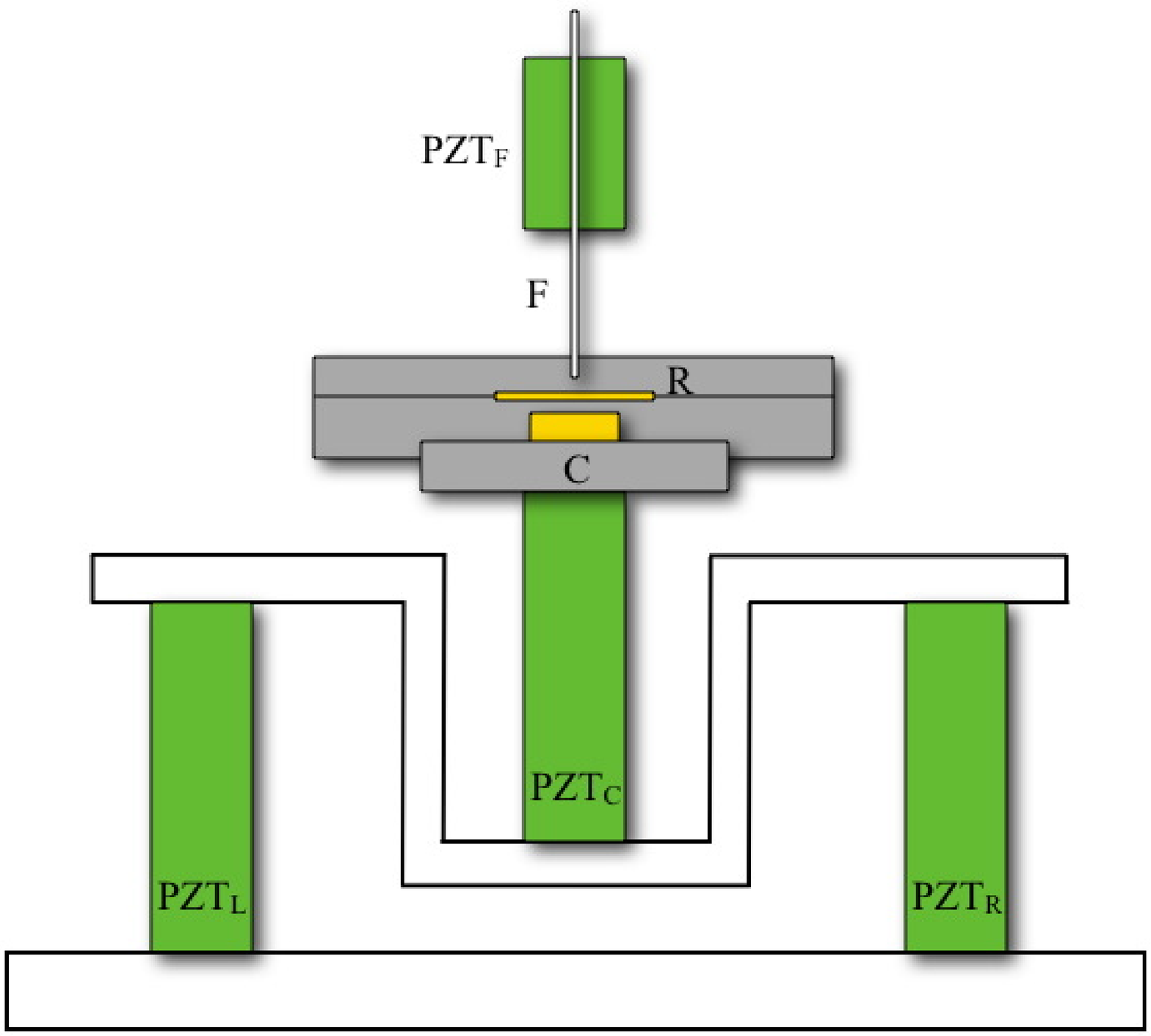} 
\includegraphics[width=0.45\columnwidth]{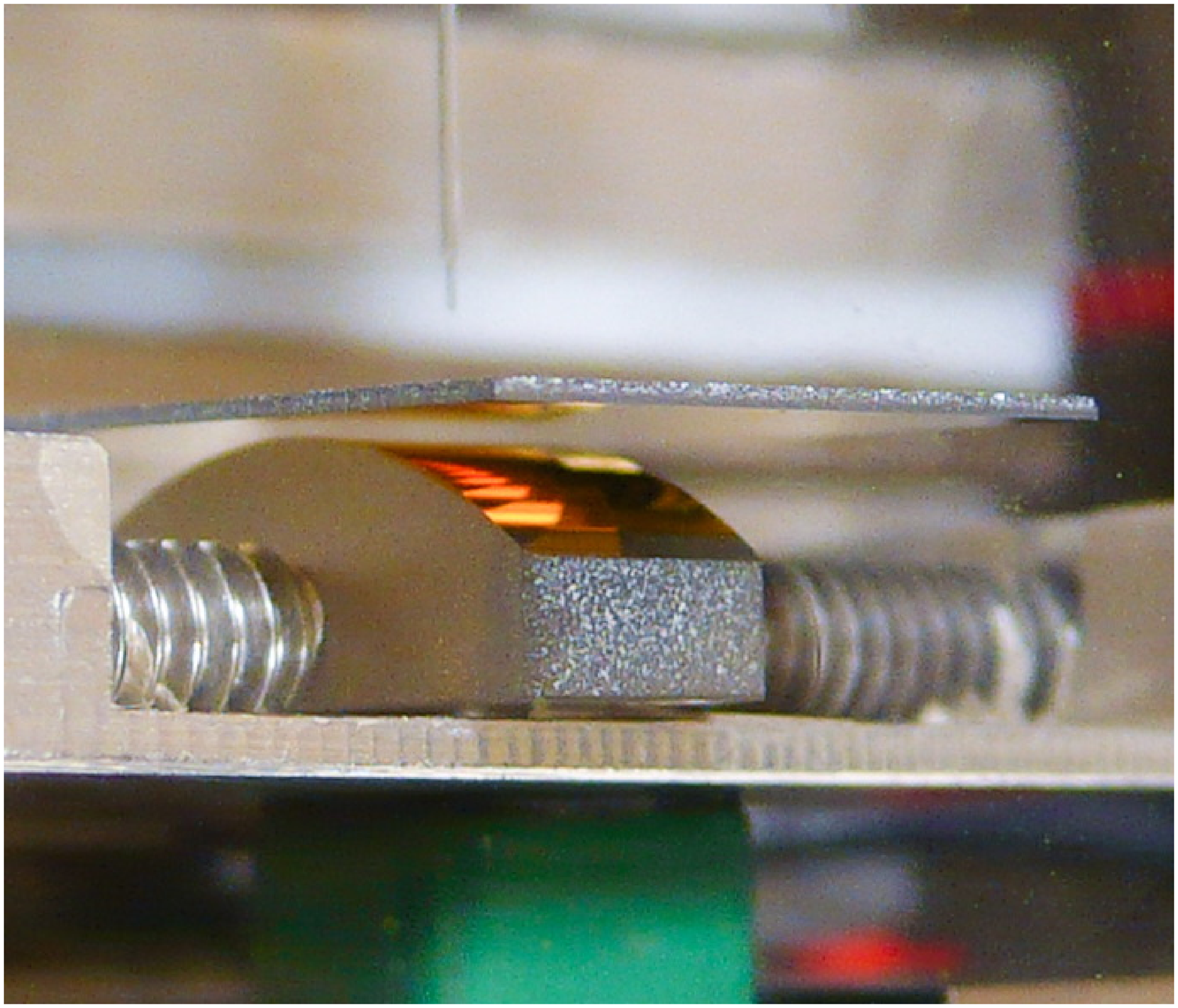}
\caption{(Color online)  
(Left) Schematic view of the experimental set-up with two lateral piezoelectric 
actuators (PZT$_\mathrm{L}$ and PZT$_\mathrm{R}$) for fine adjustment of 
parallelization, and a central piezoelectric actuator (PZT$_\mathrm{C}$) 
for controlling the cylinder-plane distance. On the top of the central 
actuator a platform is located with screws to adjust the horizontal 
position of the Au coated cylindrical lens (C). The resonator (R) is 
on the top of the cylindrical lens, and below the optical fiber (F) 
which is attached to a piezoelectric actuator (PZT$_\mathrm{F}$). 
(Right) Close-up image of the cylindrical lens and resonator region. 
The cylindrical lens has a radius of curvature of $a=12$ mm and a length 
$L=4$ mm, smaller than the width of the resonator of 10 mm.}
\label{fig1}
\end{figure}

Recently, the use of a configuration with intermediate features
between the parallel plate and the sphere-plane ones, {\it i.e.} 
the cylinder-plane geometry, has been proposed, and its experimental 
feasibility was investigated at gaps of the order of 20 $\mu$m, limited 
by the roughness of the metallic surfaces \cite{Dalvit,Michael,Michael1}. 
This geometry is very relevant from the theoretical viewpoint, since 
an exact solution for the Casimir force has been found \cite{Emig,Bordag}, also  
providing another example of curved geometry in which PFA may be tested 
against numerical techniques \cite{Gieslangfeld,Giesking}.
In this paper we report on the results of electrostatic calibrations 
for an apparatus using the cylinder-plane geometry in a range of 
distances of relevance for measuring the Casimir force. 
We observe a background originating frequency shifts of amplitude large 
enough to overwhelm the downshift expected from the Casimir force. 
We also discuss the distance dependence of the contact potential in both 
the cylinder-plane and parallel plates configurations. 
The minimizing potential shows no significant distance dependence 
in the parallel plates configuration with respect to corresponding cases 
of the cylinder-plane and sphere-plane configurations. 

\begin{figure}[ht]
\centering
\includegraphics[width=0.5\columnwidth, clip=true]{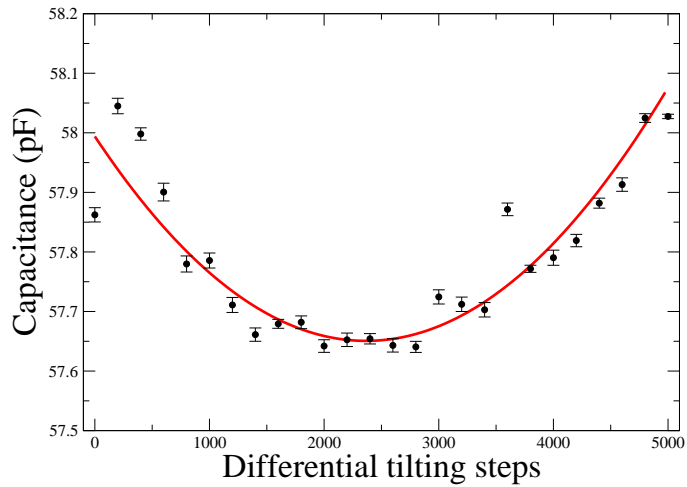}
\caption{(Color online) Assessment of the parallelization through capacitance
measurements. The cylinder-plane capacitance is shown vs. the 
difference between the steps traveled by the left and right actuators 
in a differential mode. Each data point  is the average of 20 measurements 
with an integration time of 1 s. The error bar represents the mean 
standard deviation on each data set.}
\label{fig2}
\end{figure}

The paper is organized as follows: in Sec, II we briefly recall the
Coulomb force in the cylinder-plane geometry, and report on the
upgrades to the apparatus with respect to the one described in
\cite{Michael}, its overall sensitivity performance, its geometrical 
characterization, and the parallelization technique.
In Sec. III we describe results from electrostatic calibrations, 
showing 
that in analogy to the sphere-plane case we observe that (a) the optimal exponent 
for fitting the dependence of the Coulomb coupling on distance is not the 
one expected from the idealized situation, at least at the smallest 
explored distances between the cylinder and the plane and (b) 
that the minimizing potential depends on distance. 
We then describe the data analysis leading to force residuals after 
subtraction of the Coulombian contribution.
At the smallest explored gaps we observe residual 
frequency shifts of amplitude large enough to prevent 
the measurement of the Casimir force.  
The presence of shifts neither of Coulomb nor of Casimir origin has been confirmed by
implementing a measurement strategy consisting in progressively approaching the two 
surfaces at constant bias voltage, and measuring the resonator frequency.
In Sec. IV we discuss possible explanations for the unexpected scaling 
law of the Coulomb interaction with distance. 
We then describe in Sec. V electrostatic calibrations taken in a parallel 
plane geometry aimed at evidencing the dependence on distance of the 
minimizing potential also in this configuration, thereby completing the 
analysis for the three most common geometries of experimental interest.
The relevance of measuring and modeling the dependence of the 
minimizing potential on distance in order to detect the thermal 
contribution to the Casimir force is discussed in Sec. VI.  
In the conclusion we put our findings in the more general framework
of the recent observations of systematic effects highlighting 
possible future developments for the cylinder-plane geometry.

\section{Cylinder-Plane Configuration: geometrical considerations}

The theory related to the cylinder-plane geometry and a description 
of the apparatus have been the subject of a former paper
\cite{Michael}, while further details of the measurement technique 
have been reported in \cite{Michael1,Moriond,PRARC,JPCS}. 
The main difference from the former tests is due to the use of 
high-quality Au-coated cylindrical lenses, with planarity 
and roughness comparable to the ones of the resonator,
therefore allowing us to reach submicrometer gaps.

The electrostatic force between a conducting cylinder (of length $L$, 
radius $a$, with $L \gg a$) parallel to a conducting planar surface, 
separated by a gap $d$ and kept at a fixed electrostatic potential
difference $V$ is \cite{Smythe}

\begin{equation}
F_{\rm El}^{(0)} = \frac{ 4 \pi \epsilon_0 L V^2}{\Delta \ln^2 
\left( \frac{h-\Delta}{h+\Delta} \right)} \approx
\frac{\pi \epsilon_0 \sqrt{a} L V^2}{2 
\sqrt{2} d^{3/2}}, 
\label{forceSmythe}
\end{equation}
where $\Delta=\sqrt{h^2-a^2}$ and $h=d+a$, with the approximate expression 
in the right-hand side (r.h.s.) valid in the limit $d \ll a$, also coinciding 
with the result expected from the PFA for electrostatics. 
Since both the cantilever and the cylinder have finite size, the 
length $L$ in the previous expressions should be replaced by an effective 
length $L_{\rm eff}$ that characterizes  the relative exposure between 
the cantilever and the lens, ({\it i.e.} the minimum between the width 
of the cantilever and the length of the cylinder). In our initial 
experimental attempts this corresponded to the cantilever width. 
However, by visual and optical microscope analysis we noticed the presence of sharp irregularities 
at the border of the cantilever, most likely originated by the laser cutting process of 
the Si wafer. We have then chosen cylindrical lenses of length $L=4$ mm
smaller than the resonator width of 10 mm (see Fig. 1), as the lenses seem to have more 
regular borders, as visible at the optical microscope. At this point we should note that
the curvature of our cylindrical lens is $a=12$ mm, larger than its length $L$. This implies that 
border effects in the electrostatic interaction between the cylinder and the cantilever can be important,
and the exact logarithmic expression of the electrostatic force in Eq.(\ref{forceSmythe}) should not
hold in our configuration. However, as long as the conditions for the PFA hold ($d \ll a$) the approximate
expression for the force as given by the r.h.s. in Eq.(\ref{forceSmythe}) should apply, irrespective of the 
relative magnitude of $L$ and $a$. Other important geometrical issues are the possible nonperfect
parallelization between the cylinder and the cantilever and the 
use of a cylindrical lens rather than a full cylinder.
In \cite{Michael}, the correction to the PFA expression for 
the force [r.h.s. of Eq. (\ref{forceSmythe})] due to nonparallelism was computed

\begin{equation}
F_{\rm El}^{\rm np} = F_{\rm El}^{(0)}  \frac{1}{\alpha}
\left( \frac{1}{\sqrt{1-\alpha}} - \frac{1}{\sqrt{1+\alpha}} \right) 
\approx F_{\rm El}^{(0)} \left[ 1 + \frac{5}{8} \alpha^2 + O(\alpha^4) \right],
\label{emnonparallel}
\end{equation}
where $\alpha= L \sin \theta/ 2 d$, $\theta$ is the deviation
angle from ideal parallelism, and in this nonparallel case the distance $d$ 
between the cantilever and the cylinder is measured from the midpoint along the 
axis of the cylinder. The use of a cylindrical lens rather than a full cylinder 
can be simply evaluated in PFA \cite{Moriond}, resulting in a subleading PFA 
correction (of the order of $5 \times 10^{-2}$ for a typical separation
of $d/a=10^{-3}$), and will be then discarded in what follows.

To calibrate the apparatus, a controllable electrostatic force is 
generated by applying bias voltages between the cantilever and the cylindrical lens.  
At a given separation, the frequency of a resonant mode 
characterized by an effective mass $m_\mathrm{eff}$ is measured  
both with ($\nu$) and without ($\nu_0$)  the presence of the  voltage. 
This allows for the evaluation of the square frequency difference 
$\Delta\nu^2=\nu^2-\nu_0^2$, related to the voltage $V$ in the PFA ($d \ll a$) as

\begin{equation}
\Delta\nu_{\rm el}^2 = - \frac{3\epsilon_{0}\sqrt{a}L_\mathrm{eff}}
{16\sqrt{2} \pi m_\mathrm{eff}} \frac{(V-V_0)^2}{d^{5/2}}
\left[
\frac{\alpha^{-1}}{3 (1-\alpha)^{3/2}} - \frac{\alpha^{-1}}{3 (1+\alpha)^{3/2}} 
\right] ,
\label{cpelectro}
\end{equation}
where $V_0$ is the minimizing potential. For small tilting angles $\alpha \ll 1$ 
the correction to the frequency shift due to nonparallelism has
a quadratic dependence on $\alpha$, given as $1+ 35 \alpha^2/24$.
The parallelization procedure is one-dimensional, thereby simpler 
than in the case of two flat surfaces. 
Rather than measuring the frequency shift induced by a constant 
bias voltage as discussed in \cite{Michael}, we have opted to monitor 
the cylinder-plane capacitance using a capacitive AC bridge \cite{Moriond}. 
Using the PFA, the capacitance between the cylinder and the plane is given by

\begin{equation}
C= \frac{2 \pi \epsilon_0 L_\mathrm{eff} \sqrt{a}}{\sqrt{2 d}} 
\frac{\left( \sqrt{1+\alpha} - \sqrt{1-\alpha} \right)}{\alpha}.
\label{capacitance}
\end{equation}
The tilting angle $\theta$ (and therefore $\alpha$ in turn) is
controlled through two motorized actuators acting in differential
mode, with the goal to keep the average distance $d$ (the separation 
between the plate and the cylinder as measured from the midpoint of
the cylinder) constant while changing the tilting angle (see Fig. 1).
The precision of the achieved parallelism is then determined by the 
quality of the fitting of the capacitance versus differential steps 
number, propagated to determine the dispersion on the number of steps 
at minima compatible with the fitting error. 

The precision can be improved by minimizing stray capacitance between various 
contacts since the precision of the capacitance meter is some percentage 
of the total capacitance in the system, usually $0.05\%$. An example 
of parallelization through capacitance measurements is shown in 
Fig. \ref{fig2}. The use of long-range actuators has been 
shown to be problematic due to large hysteresis, and 
therefore we have opted for a fine tuning with the use of two 
piezoelectric actuators as in the schematics shown in the left plot of Fig. 1. 
An optical microscope allows to monitor the quality and cleanness of the two 
surfaces and for a visual, qualitative assessment of the parallelization. 
An {\it a posteriori}, off-line check is obtained by fitting the electrostatic 
curves of the frequency shift using Eq. (\ref{cpelectro}), which results 
in a value of $\theta$ compatible with zero within a fitting error 
of $\delta \theta =10^{-3}$ radians. 

\begin{figure}[b]
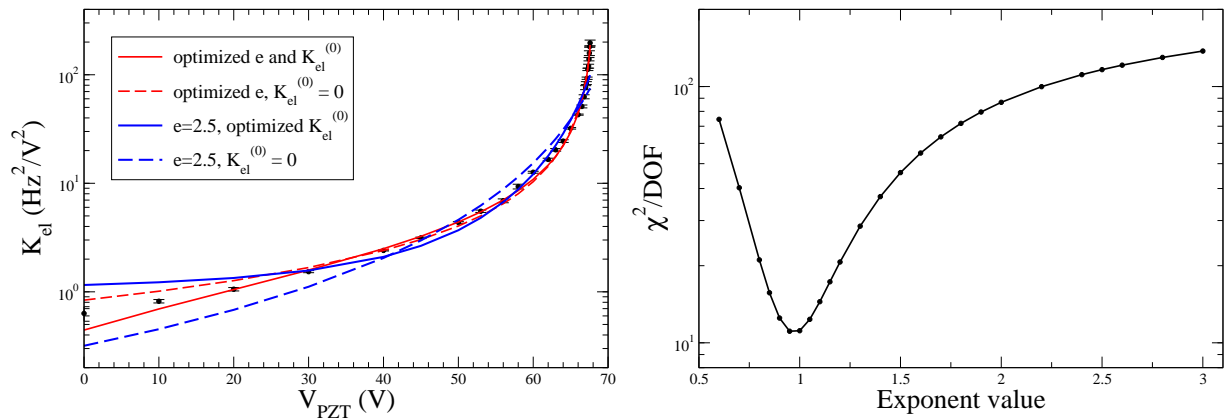

\includegraphics[width=0.45\columnwidth, clip=true]{cylindernew.fig3a.eps}
\includegraphics[width=0.45\columnwidth, clip=true]{cylindernew.fig3b.eps}
\caption{(Color online) (Left) Plots of curvature coefficient $K_\mathrm{el}$ vs. $V_\mathrm{PZT}$ 
in cylindrical-plane electrostatic measurements obtained with the curvature technique, and best fits with
the expected Coulombian interaction having a 2.5 exponent (blue/dark gray continuous curve), and with 
a power-law functional dependence in which the exponent is instead a free parameter (red/light gray 
continuous curve). The dashed curves are obtained by also constraining to zero an offset 
for $K_\mathrm{el}$, representing a possible curvature present even 
at a large distance between the cylinder and the plane, for instance 
due to stray environmental electric fields. 
(Right) Plot of the reduced $\chi^2$ obtained by dividing $\chi^2$ by the degrees 
of freedom (DOF) vs. the value of the power
exponent in the case of the offset $K_\mathrm{el}^{(0)}=0$.
Data are taken as described in \cite{PRARC,JPCS}, with a modification of the 
acquisition code for faster data acquisition and smaller uncertainty from drift effects. 
In the routine used for data acquisition in \cite{PRARC,JPCS}, the bias voltage always 
returns to 0 V after each measurement. 
In the new routine instead, it changes from $+V_b$ to $-V_b$ in steps of 
$\delta V$ where $V_b$ and $\delta V$ ({\it i.e.} the maximum bias voltage and 
its minimum step of variation during the calibration) are specified in advance in the code. 
This allows to double the acquisition speed for a targeted span of voltage values. 
Although this sacrifices some accuracies in
the subsequent fitting of the data, a shorter data acquisition time 
is preferable considering the relatively large drift experienced 
during the entire duration of a typical run.}
\label{fig:optimalexponent}
\end{figure}

\section{Electrostatic calibrations and residuals analysis in the cylinder-plane geometry}

An important prerequisite to any Casimir force measurement is the
execution of high-quality electrostatic calibrations. 
Forces are always indirectly measured via their functional 
relationship to more accessible observables, such as detection 
of deflection angles, voltages required to keep the apparatus at 
rest in closed loop schemes, or shifts of the frequency of a mode 
of a resonator as in our case. It is then crucial to convert 
such quantities more directly observed into the corresponding force signal 
by means of well-known and controllable physical signals, for instance 
by applying external bias voltages and comparing the measured forces 
with the Coulombian interactions between macroscopic conducting bodies.  

The electrostatic calibration starts by finding the best parallelization 
condition at a given nominal cylinder-plane separation $d$ using the 
capacitance technique described previously. Then the parallelization 
is further fine tuned by adjusting the two lateral piezoelectric 
actuators ${\rm PZT}_{\rm L,R}$. Once the optimal parallel
condition is obtained, ${\rm PZT}_{\rm L,R}$ are left untouched,
and the separation $d$ is changed via the central piezoelectric 
actuator ${\rm PZT}_{\rm C}$. This procedure keeps the parallelization 
at the optimal value. As seen previously, in the perfect parallel 
situation the frequency shift of the cantilever due to pure 
electrostatics takes the form

\begin{equation}
\Delta \nu_\mathrm{el}^2
= - \frac{3\epsilon_0 \sqrt{a} 
L_\mathrm{eff}}{16 \sqrt{2} \pi m_\mathrm{eff}}\frac{(V-V_0)^2}{d^{5/2}}=
-K_{\rm el}(V-V_0)^2.
\label{eqn:nu_cp}
\end{equation}
The displacement of the central piezo ${\rm PZT}_{\rm C}$ depends
linearly  on the voltage applied $V_\mathrm{PZT}$, and therefore the absolute 
gap is given by  $d=\beta(V^0_\mathrm{PZT}-V_\mathrm{PZT})$, where $\beta=(91.9 \pm 0.9)$ nm/V 
is the actuation coefficient of the piezoelectric transducer, and 
$V^0_\mathrm{PZT}$ is the PZT voltage required to make contact 
between the two surfaces. At a given distance, the electrostatic 
calibration has been performed by measuring the frequency shift 
induced by a range of electric voltages $V$ applied between the two 
surfaces. The curvature coefficient $K_{\rm el} = 3 \epsilon_0 
\sqrt{a} L_{\rm eff} / 16 \sqrt{2} \pi m_{\rm eff} d^{5/2}$ 
and the minimizing potential $V_0$ can then be obtained by fitting the data with 

\begin{eqnarray}
&&  \nu^2_{\rm el} = \nu^2_{\rm el}(V, V_{\rm PZT}) = \nu^2_{0} - K_{\rm el}(V_{\rm PZT}) 
\times  (V-V_0)^2 , \label{eqn:Kel_vs_d}\\
&& K_\mathrm{el}(V_{\rm PZT}) = \gamma (V^0_\mathrm{PZT}-V_\mathrm{PZT})^{-5/2} , 
\end{eqnarray}
where $\gamma \equiv 3 \epsilon_0 \sqrt{a} L_{\rm eff} / 16 \sqrt{2} 
\pi m_\mathrm{eff} \beta^{5/2} $.
This fitting procedure allows the determination of the absolute distance $d$ 
once the fitting parameter $V^0_\mathrm{PZT}$ is obtained, and the
measurement of the contact potential $V_0$ as a function of $d$. 
A typical data plot of $K_\mathrm{el}$ {\it vs.} $V_\mathrm{PZT}$ is shown 
in the left plot of Fig.~\ref{fig:optimalexponent}. The blue curves are the best fits
using Eq.~(\ref{eqn:Kel_vs_d}), and they deviate significantly from the 
data points, both including or excluding a curvature offset 
$K_\mathrm{el}^{(0)}$ representing a hypothetical background electric field.
Moreover, the effective mass calculated from the fitting parameter 
is 30-50 times larger than the physical mass, much larger than the  
expected value of the effective mass which should be comparable or smaller than the physical 
mass of the resonator. If instead the power exponent is left as a free 
parameter, rather than being fixed at 2.5, a new fitting curve with 
the exponent in the 0.9 to 1.3 range is obtained (the red curves 
in Fig.~\ref{fig:optimalexponent}). This deviation from the expected 
exponent for the Coulombian force has been confirmed to exist in all 
our electrostatic calibrations data.

Table I shows the fitting parameters of the electrostatic calibrations for five runs both
when the exponent is fixed and left  as a free parameter. The effective mass for the latter 
case is not well defined because of the deviation of the exponent  from 2.5, thus it is not listed.
As shown in Table I, the exponent, when left as a free parameter, is
always smaller than the theoretical value of 2.5, with a reduced 
$\chi^2$ smaller by about an order of magnitude with respect to the 
one expected from the Coulombian scaling. The relatively large value of
$\chi^2$ also indicates that the errors may be underestimated, although 
this does not affect our conclusions about the relative comparison 
between Coulombian and optimal exponents.
While small deviations from 2.5 are expected considering all the less than 
ideal conditions such as imperfect parallelization and thermal and mechanical drifts, 
such a significant difference (an average value of 1.03 
versus 2.5) cannot be explained as small deviations from ideality. 

\begin{table}[t]
\begin{center}
\begin{tabular}{|l|c|c|c|c|c|}
       \hline \hline
                          & Run 1             &  Run 2            &   Run 3            &   Run 4            &   Run 5  \\
       \hline
fixed exponent            &  2.5              &   2.5             &    2.5             &    2.5             &    2.5      \\
$V^0_\mathrm{PZT}$ (V)    &  79.52$\pm$0.06   &  73.08$\pm$0.06   &   56.76$\pm$0.03   &   71.05$\pm$0.02   &   65.84$\pm$0.02   \\
$d_\mathrm{min}$ (nm)     &  590              &  504              &   491              &   464              &   477    \\
$m_\mathrm{eff}$ (g)      &  6.63             &  9.25             &   8.87             &   9.46             &   9.01    \\
$\chi^2$/DOF              &  60               &  116              &   283              &   518              &   876     \\
       \hline
free exponent             &   1.30$\pm$0.03   &   0.97$\pm$0.01   &    0.93$\pm$0.01   &    1.00$\pm$0.01   &    0.94$\pm$0.01  \\
$V^0_\mathrm{PZT}$ (V)    &  74.37$\pm$0.04   &  68.07$\pm$0.02   &   51.78$\pm$0.01   &   66.45$\pm$0.01   &   61.03$\pm$0.01  \\
$d_\mathrm{min}$ (nm)     &  116              &  43               &   33               &   42               &   34  \\
$\chi^2$/DOF              &  5.3              &  11               &   19               &   42               &   31   \\
\hline
\hline
\end{tabular}
\caption{Fitting parameters of the electrostatic calibrations using the curvature technique.}
\end{center}
\label{tab:tableexponent}
\end{table}

\begin{figure}[b]
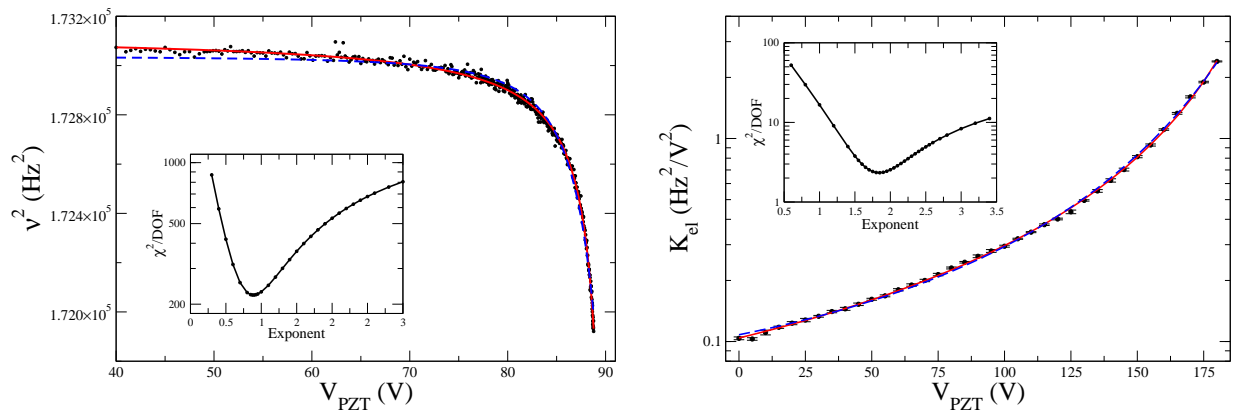

\includegraphics[width=0.45\columnwidth, clip=true]{cylindernew.fig4a.eps}
\includegraphics[width=0.45\columnwidth, clip=true]{cylindernew.fig4b.eps}
\caption{(Color online) (Left) Squared resonance frequency versus 
$V_\mathrm{PZT}$ for the fast-approach calibration technique. 
The data were obtained using a measurement protocol in which 
a constant bias voltage equal to 4 V was applied across the 
two surfaces and the resonance frequency was progressively 
measured from the farthest to the closest distances. 
The dashed blue curve is the fitting with a 2.5 exponent while 
the continuous red curve is obtained leaving the exponent as a free 
parameter, whose optimal value turned out to be about 0.89. 
In the inset the reduced $\chi^2$ is plotted 
versus the value of the free exponent. (Right)  Test of the electrostatic scaling law for the 
cylinder-plane geometry at large distances with the curvature 
technique. Plot of curvature coefficient $K_\mathrm{el}$ versus 
$V_\mathrm{PZT}$. The red curve is the fit with 2.5 exponent, the 
blue is the fit with the optimal exponent 1.84. 
In the inset the reduced $\chi^2$ is plotted versus the value of 
the free exponent, with a minimum of $\chi^2$ obtained for an exponent 
equal to 1.84.}
\label{fig:fastfarexponent}
\end{figure}

Electrostatic calibrations have also been performed with an alternative technique 
consisting in directly measuring the resonance frequency as the separation gap is 
decreased maintaining a constant bias voltage. 
This so-called {\sl fast-approach} measurement technique has 
the advantage of a faster data acquisition, resulting in a mitigation 
of the long-term thermal or mechanical drifts, and provides an alternative 
to check the distance dependence of the electrostatic force. 
As shown in the left plot of Fig.~\ref{fig:fastfarexponent}, the optimal exponent obtained from 
the minimum of the reduced $\chi^2$ curve is around 0.89. 
Although this technique seems to be slightly less sensitive to the 
exponent, as shown from the softer dependence of $\chi^2$ (the 
value of $\chi^2$ for the exponent of 2.5 being only three times larger 
than the minimum $\chi^2$ obtained at the exponent of 0.89), the 
optimal value of the exponent is consistent with the results from the 
previous technique based upon electrostatic calibrations. 
The fact that a significant deviation of the exponent from 2.5 is still 
observed even in {\sl fast-approach} measurements 
limits or rules out the possibility that systematic effects, such 
as artifacts from fitting the parabolic dependence in Eq.~(\ref{eqn:nu_cp}), 
or long-term drifts such as thermal expansions or relaxation of the 
PZT actuator, may be responsible for this anomalous behavior. 

As can be seen in Table I, the smallest distance achieved is of the
order of 500 nm based on the fitting with a fixed exponent of 2.5. 
In a previous measurement using wider cylindrical lenses with a larger 
radius of curvature the smallest distance reached was around
1 $\mu$m, and no significant deviation of power exponent from 2.5 has
been observed. This suggests that the exceptionally small exponent may be a
result of the smaller gaps reachable with the cylindrical lenses of 
smaller width and radius of curvature. 
Although the absolute distances obtained from the fitting with
exponent 2.5 cannot be fully trusted in light of the relatively
inaccurate fitting, they can be still considered as reliable enough 
to estimate the gap separation. 

\begin{figure}[t]
\includegraphics[width=0.5\columnwidth, clip=true]{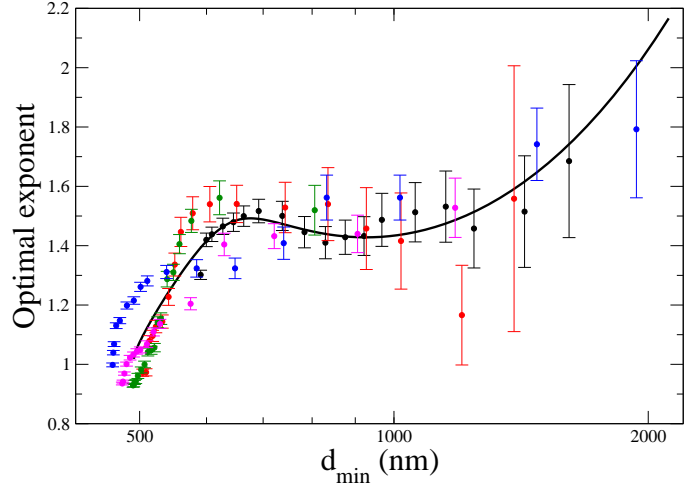}
\caption{(Color online) Optimal exponents of five electrostatic calibrations 
versus the distance of the closest data point used in the fitting, with the 
continuous line representing a spline curve obtained by considering all data 
points, weighted by their error bars.}
\label{fig:allexponent}
\end{figure}

\begin{figure}[b]
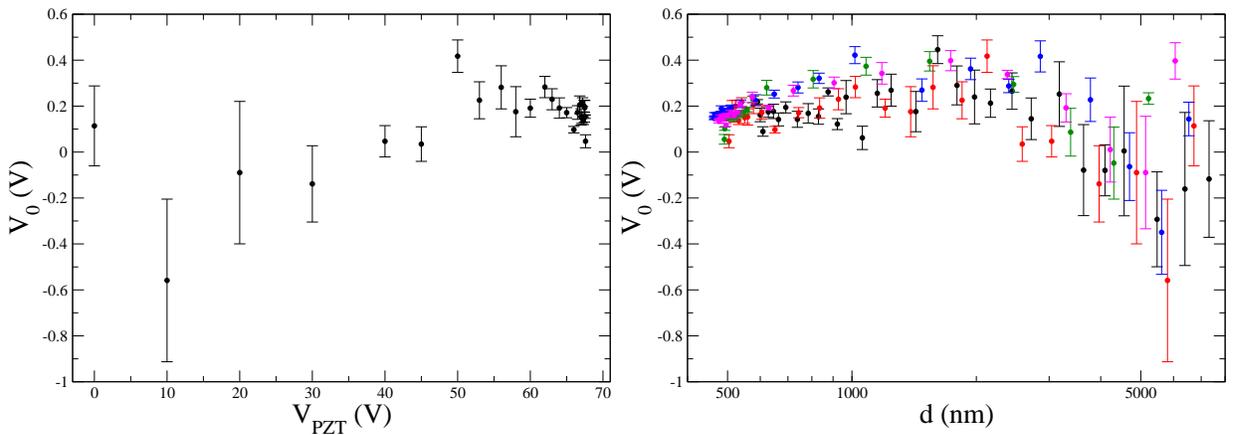

\includegraphics[width=0.45\columnwidth, clip=true]{cylindernew.fig6a.eps}
\includegraphics[width=0.45\columnwidth, clip=true]{cylindernew.fig6b.eps}
\caption{(Color online) Dependence of the minimizing potential on distance in the
  cylinder-plane geometry. (Left) Minimizing potential $V_0$ {\it versus} 
$V_\mathrm{PZT}$ from a typical electrostatic calibration measurement.  
(Right) Minimizing potential $V_0$ versus distance $d$ from various runs.}
\label{fig:residualv}
\end{figure}

Electrostatic calibrations with the same lens were performed at
relatively large distances, as shown in the right plot of Fig. 4. 
In the data presented in Table I, the explored distances ranged 
from about 500 nm to 7.3 $\mu$m. The new calibrations were instead 
performed in the range of about 5.1 to 21.6 $\mu$m. 
This obviously results in much smaller frequency shifts and 
larger error bars in the values of $K_\mathrm{el}$, yet 
it is evident from the fitting that an exponent of 2.5 is 
more satisfactory for these large-distance data. 
Furthermore, the fitting parameters with an exponent fixed at 2.5 
gives an effective mass of 1.2 g which is much closer to the 
estimated physical mass of the resonator of 0.2 g. 
If the power exponent is instead left as a free parameter, 
a value of 1.84$\pm$0.06 is obtained. The large-distance data span 
smaller ranges of $K_\mathrm{el}$ and therefore are less sensitive to the
power exponent and as shown in the inset on the left plot of 
Fig.~\ref{fig:fastfarexponent}, the difference between the 
fitting curve of 2.5 exponent and that of 1.84 exponent is 
not very significant (corresponding to a reduced $\chi^2$ of 
2.3 versus 4.9). Nevertheless, the fitting makes evident that the optimal
exponent at large distance is almost two times larger than the 
one obtained from the small-distance data. 
We also investigated how the optimal exponent changes 
if a subset of data, rather than the entire set, is used 
in the fitting procedure. In particular, we have fitted 
subsets of data obtained by progressively removing points 
at the smallest distances. As shown in Fig.~\ref{fig:allexponent}, 
all five runs share the same trend showing that the optimal 
exponent  increases when the number of the removed point 
of closest distance used in the fitting increases. 
The absolute distances in the plot were
obtained from the fitting with fixed 2.5 exponent. 
One feature which is noticeable in the figure is a relatively sharp increase 
in the value of the optimal exponent for a distance range of 500 to 600 nm. 
This could explain why such a large deviation of the optimal exponent from 2.5 
has not been observed in our earlier measurements with a larger radius of curvature 
cylinders in which we managed to reach minimum distances of only about 1 $\mu$m.
We have also noticed a strong correlation between the value of the effective 
mass $m_\mathrm{eff}$ and the exponent. The effective mass obtained from the 
fitting increases when the distance between the two surfaces decreases. 
The dependence on distance of the optimal exponent is in contrast to the case of 
the sphere-plane measurements in which a relatively constant optimal exponent 
was observed uniformly over the entire range of explored distance \cite{PRARC,JPCS}.
All fits are performed with a weight equal to $1/\sigma_i^2$, where 
$\sigma_i$ is the standard deviation of $K_\mathrm{el}$ from the
parabola fitting. 

In our experiment we have also studied the possible distance variability of the
minimizing potential $V_0$ ({\it i.e.} the voltage difference between 
the cylindrical and the planar conducting surfaces which is minimizing 
the electrostatic force, as described for instance in \cite{PRARC,JPCS}). 
At larger gaps we have observed an approximate linear relationship 
between the residual potential $V_0$ and $V_\mathrm{PZT}$ in the 
cylinder-plane configuration. With closer approaching (larger values of $V_{\rm PZT}$),  
$V_0$ tends to have a nearly flat dependence on the distance, 
as is visible in Fig. \ref{fig:residualv}, both on a single run (left plot) and on various 
runs obtained in different days (right plot).

\begin{figure}[t]
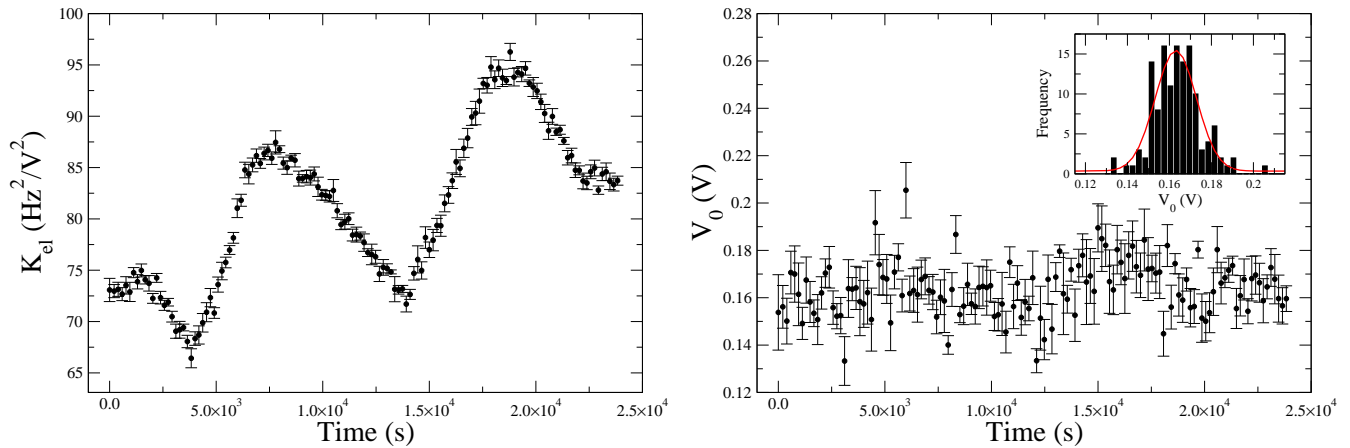

\includegraphics[width=0.49\columnwidth, clip=true]{cylindernew.fig7a.eps}
\includegraphics[width=0.49\columnwidth, clip=true]{cylindernew.fig7b.eps}
\caption{Plots of the curvature coefficient $K_\mathrm{el}$ (left) and  
the minimizing potential $V_0$ (right) versus time at constant $V_\mathrm{PZT}$. 
The inset of the right plot is the histogram of $V_0$ which shows that $V_0$ 
roughly follows a Gaussian distribution centered around 0.163 V with a 
full width at half maximum of 0.023 V (standard deviation 0.01 V).}
\label{fig:drifts}
\end{figure}

To estimate the effect of long-term drifts we have studied the
time dependence of the electrostatic curvature coefficient $K_\mathrm{el}$ 
and the minimizing potential $V_0$, without nominally changing the 
cylinder-plane gap distance, as shown in Fig.~\ref{fig:drifts}. 
The curvature coefficient shows temporal variations of order 
50 $\%$, while the minimizing potential $V_0$ does not show any 
evident dependency, rather it fluctuates in a relatively small range of values. 

\begin{table}[b]
\begin{center}
\begin{tabular}{|l|c|c|c|c|c|}
       \hline \hline
$d_\mathrm{min} - d_\mathrm{touch}$ (nm) &  0             & 92            & 184           &  276          &  368 \\
       \hline
$V_\mathrm{bias}$ = 3V                   & -0.91$\pm$0.02 & -1.52$\pm$0.10 & -1.56$\pm$0.17 & -1.91$\pm$0.32 & -2.05$\pm$0.51 \\
$V_\mathrm{bias}$ = 4V                   & -0.89$\pm$0.03 & -1.20$\pm$0.08 & -1.73$\pm$0.19 & -1.89$\pm$0.30 & -1.83$\pm$0.36 \\
$V_\mathrm{bias}$ = 5V                   & -0.98$\pm$0.02 & -1.70$\pm$0.04 & -1.60$\pm$0.06 & -1.61$\pm$0.10 & -1.62$\pm$0.16 \\
\hline
\hline
\end{tabular}
\caption{Optimal exponents for various values of minimum distance used in the fitting of 
the data from fast approach measurements.}
\end{center}
\label{tab:tablefastexponent}
\end{table}

\begin{figure}[t]
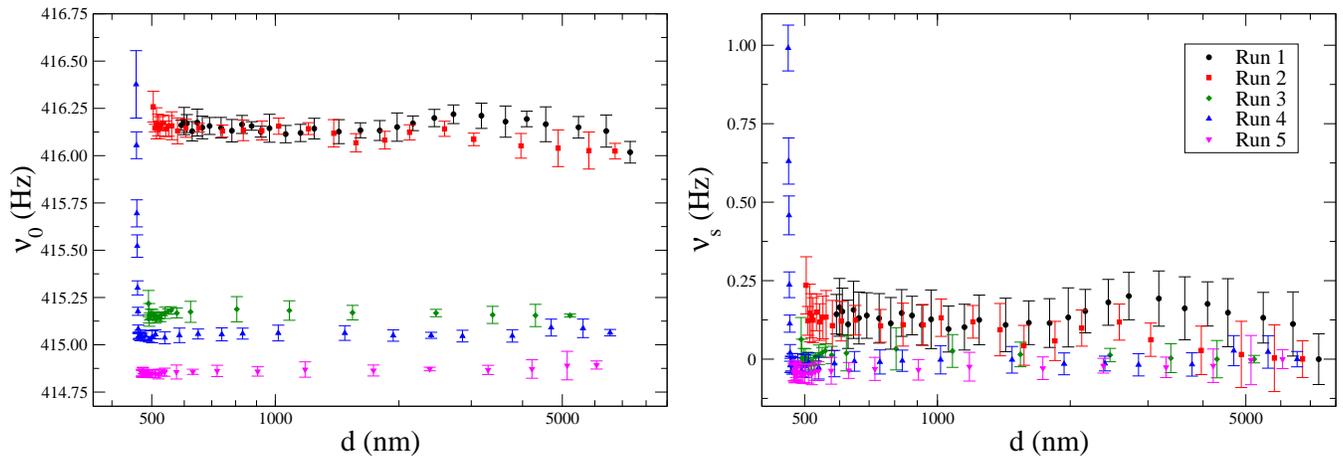

\includegraphics[width=0.49\columnwidth, clip=true]{cylindernew.fig8a.eps}
\includegraphics[width=0.49\columnwidth, clip=true]{cylindernew.fig8b.eps}
\caption{(Color online) Residuals from electrostatic calibrations with parabola method. 
(Left) Plot of residual frequencies from various runs of electrostatic calibration 
measurements {\it vs.} distance. Because of the dependence of the resonant frequency 
on temperature, different runs have different frequencies even at the largest gaps 
at which no residual force is expected. (Right) Same plot but with a common baseline 
chosen in such a way that the data points at the farthest distance have a common
central value, with $\nu_s=\nu_0-\nu_0(d_\mathrm{max})$.
}
\label{fig:nu0_cp}
\end{figure}

\begin{figure}[b]
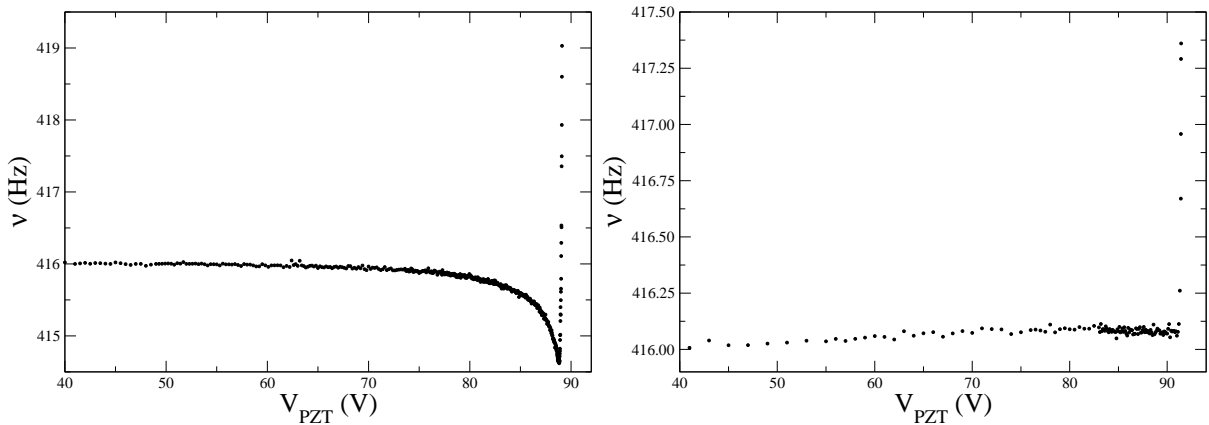

\includegraphics[width=0.45\columnwidth, clip=true]{cylindernew.fig9a.eps}
\includegraphics[width=0.45\columnwidth, clip=true]{cylindernew.fig9b.eps}
\caption{(Left) Plot of frequency versus $V_\mathrm{PZT}$ with a constant bias voltage 
$V_\mathrm{bias}$ = 4 V obtained using the fast-approach method. (Right) Plot of frequency 
versus $V_\mathrm{PZT}$ with a constant bias voltage 
$V_\mathrm{bias}$ = 0.15 V approximately compensating the value of the minimizing 
potential expected at the smaller gaps. No evidence for downshifts attributable 
to charge-independent forces is visible until the frequency increases due to 
direct contact between the two surfaces.}
\label{fig:fast4vraw}
\end{figure}

\begin{figure}[t]
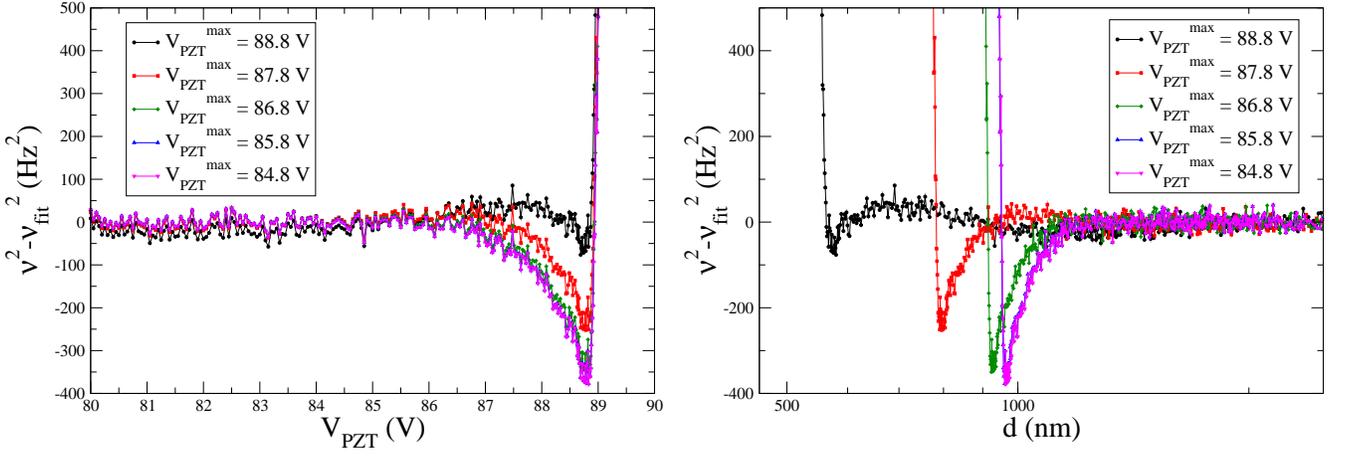

\includegraphics[width=0.49\columnwidth, clip=true]{cylindernew.fig10a.eps}
\includegraphics[width=0.49\columnwidth, clip=true]{cylindernew.fig10b.eps}
\caption{(Color online) Residuals to the frequency shifts with the fast-approach technique. 
(Left) Plots of residual frequency square ($\nu^2$ - $\nu_\mathrm{fit}^2$) 
for a constant bias voltage $V_\mathrm{bias}$=  4 V versus $V_\mathrm{PZT}$ and 
different values of the maximum value of $V_\mathrm{PZT}$ used for the data analysis. 
(Right) Plot of the corresponding force versus distance. 
Both the absolute force and the absolute distance are inferred by the 
Coulomb fitting at larger distances.}
\label{fig:fast4vresidual}
\end{figure}

\begin{figure}[b]
\includegraphics[width=0.5\columnwidth, clip=true]{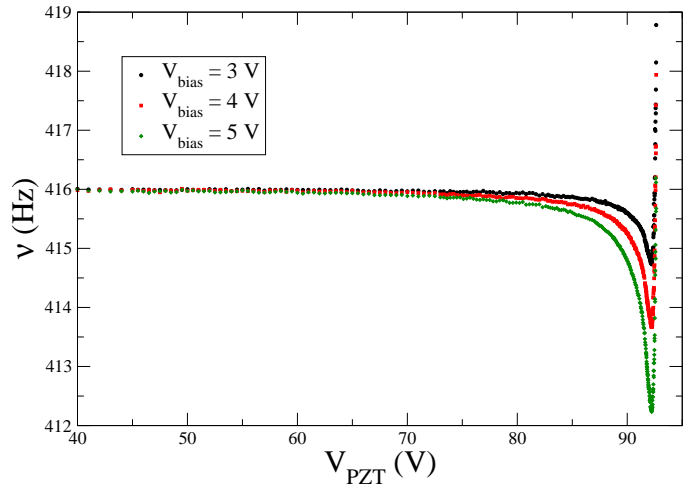}
\caption{(Color online) Fast-approach measurements with various values of the bias
  voltage. Plot of the resonator frequency versus distance with bias 
voltage $V_\mathrm{bias}$ of 3 V (black circle points), 4 V (red square points), and 5 V 
(green diamond points).}
\label{fig:3v4v5vraw}
\end{figure}

\begin{figure}[t]
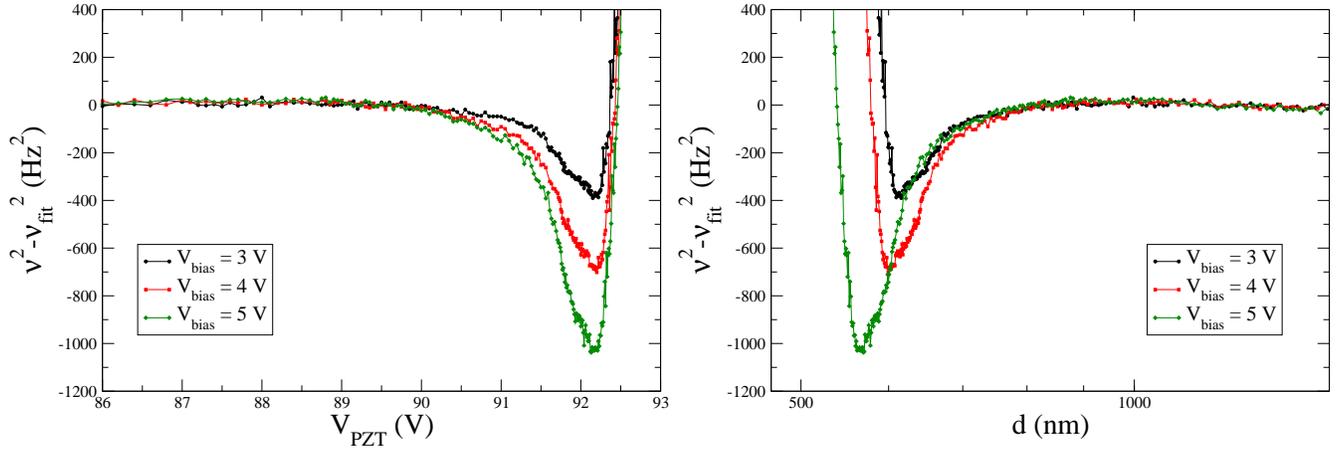

\includegraphics[width=0.49\columnwidth, clip=true]{cylindernew.fig12a.eps}
\includegraphics[width=0.49\columnwidth, clip=true]{cylindernew.fig12b.eps}
\caption{(Color online) Plots of residual frequency square ($\nu^2$ - $\nu_\mathrm{fit}^2$) from the fast-approach 
measurement with $V_\mathrm{bias}$ = 3 V (black circle points), 4 V (red square points) and 5 V (green diamond points) 
versus $V_\mathrm{PZT}$ (left) and distance (right). The fittings were done after removing the closest 184 nm data. }
\label{fig:3v4v5vresidual}
\end{figure}

\begin{figure}[b]
\includegraphics[width=0.49\columnwidth, clip=true]{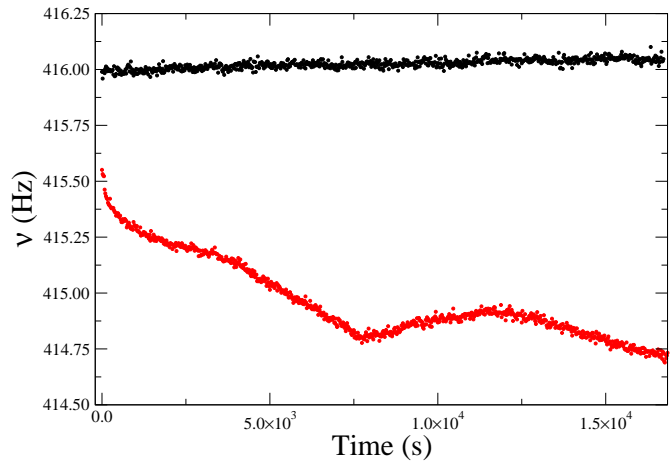}
\caption{(Color online) Plots of the resonator frequency versus time in the case of large 
separation (black upper points), and small separation (red lower points), the latter obtained by 
imposing an external bias voltage of 4 V at a nominally constant distance.
While the first plot reflects the intrinsic resonator change, for instance 
due to temperature drifts or internal creeps, the second also includes the 
effect of drifts over time in the cylinder-plane separation distance. }
\label{fig:frequencydrifts}
\end{figure}

From the electrostatic calibration measurements it is also possible 
to extract the electrostatic-free frequency $\nu_0$ as can be seen 
from Eq.~(\ref{eqn:Kel_vs_d}). In principle, when the distance between 
the two surfaces is small enough, one should expect a downward shift 
of this residual frequency  due to attractive Casimir forces. 
However, as can be seen in Fig.~\ref{fig:nu0_cp}, no significant 
downshift of $\nu_0$ has been observed, rather a sharp upshift 
was instead observed in one case. 
This upshift indicates a very strong repulsive force between the two 
surfaces at small distances, which could be from either a repulsive 
component of the Van der Waals forces or simply a soft contact of the surfaces. 
To better understand the short-distance behavior and to reduce the effect 
of thermal drifts, we have performed a series of measurements using 
the fast-approach technique mentioned previously.

As shown in Fig.~\ref{fig:fast4vraw}, with a constant bias voltage 
$V_\mathrm{bias}$=4 V, the frequency of the resonator decreases as 
the cylindrical lens approaches the resonator, as qualitatively 
expected for the attractive Coulomb force, then followed by a sharp 
increase, most likely from contact of the surfaces. 
If the Coulomb force is the dominant force, then the downward part 
of the curve can be fitted using Eq.~\ref{eqn:nu_cp}. 
As mentioned earlier, a fixed exponent of 2.5 produces a marginal fit 
while the optimal exponent is around 0.9. However, we should keep in 
mind that the exponent is not supposed to be 2.5 in the first place 
if at small distances some other forces 
(either the Casimir force, patch forces, or corrections to the standard Coulomb 
interaction mentioned in the previous section) become large enough to compete 
with the Coulomb force. Based on this remark, we further analyze the data
by removing data points at the smallest distances, to test the
stability of the fit with the Coulomb force. In these conditions 
the optimal exponent increases, as shown in Table II. 
In analogy to the previous analysis on electrostatic calibration data, 
although the optimal exponent is still smaller than 2.5, the value is 
approaching 2.5 within the relatively large error bars. This indicates 
once again that the sources of discrepancy between the data and the
expected Coulomb force are localized at the smallest distances. 

We then can investigate another kind of residual of the data to 
analyze the effect of data taken at the smallest distances. 
Instead of relaxing the exponent finding its optimal value, the 
exponent is kept fixed at 2.5 when fitting the larger distance portion of the data, and the 
{\sl residuals} at small distances are evaluated. This residual analysis is shown in 
Fig.~\ref{fig:fast4vresidual} for one run, with the soft contact 
occurring around $V_\mathrm{PZT}$ = 88.8 V. If the Coulomb force 
is the only dominant force for the whole downshifted part before 
the soft contact, then the fitting curve obtained when excluding 
a small region of data prior to contact, for instance all the data 
corresponding to a PZT voltage larger than $V_\mathrm{PZT}$ = 87.8 V, 
should be able to predict the data obtained by excluding distances 
corresponding to a further Volt of $V_\mathrm{PZT}$ removed, with 
residuals centered around zero. However, by doing so there is clearly 
a nonzero downshift residual before the soft contact takes over. 
In principle this residual could come from fitting artifacts, in 
particular it could depend on the interval chosen for fitting the 
data with the Coulomb force. 
However, when more points were removed, this nonzero residual 
appears to be stabilized. 
Considering that the frequency shifts with this fast-approach 
technique may capture all possible forces acting on the resonator, 
this electrostatic residual analysis indicates that there are forces 
other than the expected contribution from the applied constant bias 
voltage that caused a further downshift in the frequency. 
To test whether the residual force is correlated to the 
external bias electric field, measurements were also performed with 
$V_\mathrm{bias}$ = 0.15 V, corresponding to the average value of 
the residual potential $V_0$ at small distances as shown in Fig.~\ref{fig:residualv}.
No noticeable downshift was observed, as seen in the right plot of Fig.~\ref{fig:fast4vraw}. 
Furthermore, to rule out possible changes of configurations (such as parallelism  
or distance drifts) in runs taken in different days, fast-approach measurements 
were performed with different $V_\mathrm{bias}$ applied within the same run, 
({\it i.e.}, at each position three values of $V_\mathrm{bias}$ were applied 
and the respective frequencies measured, as shown in Fig.~\ref{fig:3v4v5vraw} 
with bias voltages of 3, 4, and 5 V). 
This confirms that larger bias voltages result in larger frequency 
shifts at the same distance. The residual analyses were carried out 
and the results with the fitting after removing the closest 184 nm 
data are shown in Fig.~\ref{fig:3v4v5vresidual}. 
The extra downshifts were present in all three curves, with peak value 
approximately quadratic in the external bias voltage.

Although this residual force  $F_\mathrm{res}$ is attractive, it cannot 
be identified with the sought Casimir force. 
Apart from the absence of a comparable signal in the residual frequency analysis 
with the electrostatic calibration technique (Fig.~\ref{fig:nu0_cp}), $F_\mathrm{res}$ 
seems to be dependent on the applied bias voltage as shown in 
Fig.~\ref{fig:3v4v5vresidual}. Moreover, best fits 
of this residual force with power-law expressions indicate that it is required 
an exponent which is much larger than that of the Casimir force. 
The fact that this extra force depends on the applied bias voltage indicates that it may also 
be present in electrostatic calibrations and could lead to an anomalous exponent.

It should be noticed that within our statistics of runs performed with 
the fast-approach technique (about 20) we have also observed a couple 
of runs in which residuals gave rise to short-distance upshifts. 
This could be explained by the presence of anomalous distance drifts 
due to external factors such as the environmental temperature.
To assess this effect and how much it affects the amplitude of the observed 
frequency shifts in the residuals, we show in Fig.~\ref{fig:frequencydrifts}
the typical amplitude of the frequency fluctuations. 
This allows for disentangling the intrinsic drifts due to changes in the 
resonator frequency, obtained by monitoring the resonator with a distance 
from the cylinder large enough to make negligible its influence, and drifts 
due to changes in the cylinder-plane distance, with a bias voltage of 4 V 
intermediate between the two extreme values of voltages applied in the 
fast-approach measurements.

\section{Possible explanations for the anomalous exponent}

In the following section we discuss possible causes of the anomalous exponent 
obtained in our electrostatic calibrations of the cylinder-plane geometry. 
In particular, we consider edge effects, local deformations from the idealized 
geometry, electric forces with steeper distance scaling than the Coulomb force, 
and electrostatic patch effects.

\subsection{Edge effects}

\begin{figure}[t]
\includegraphics[width=0.5\columnwidth, clip=true]{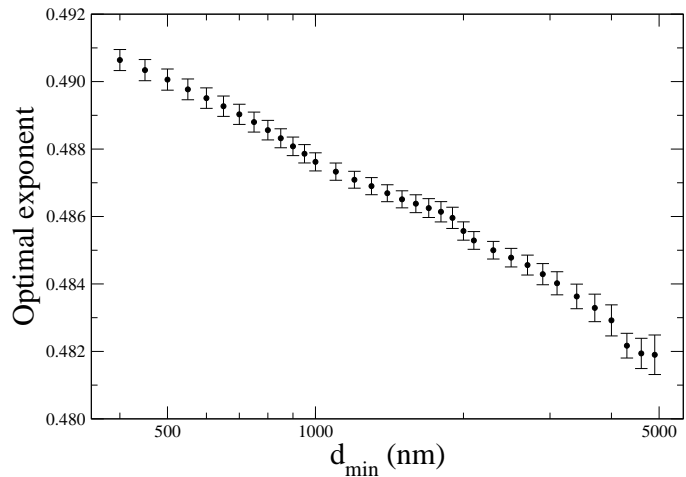}
\caption{Optimal exponent {\it vs.} the distance of the closest data point used in the fitting
of the capacitance {\it vs.} distance curve for the numerical simulation of the finite size  
cylindrical lens in front of the finite size planar resonator, the geometry corresponding 
to the actual experimental setup as shown in Fig. 1.}
\label{fig:comsolexponent}
\end{figure}

One possible reason for a deviation from the ideal Coulomb prediction of 2.5 for 
the exponent of the frequency shift versus distance is the fact that a finite 
size cylindrical lens was used in the experiment, instead of a long whole cylinder. 
In this situation edge effects may not be negligible, and Eq. (\ref{forceSmythe}) 
would not be a good approximation to the actual electrostatic frequency shift.
While edge effects have been discussed for Casimir forces with the general world-line 
approach in \cite{Gies1}, we have not found former discussions of this effect in the 
electrostatic calibrations in Casimir experiments. Using the COMSOL numerical package, 
we have conducted numerical simulations in which the precise geometry of the measurements 
was used, and the capacitance between the cylindrical lens and the resonator was 
evaluated at different distances \cite{Qun}. By neglecting edge effects, the power exponent 
for the capacitance versus distance would be 0.5. We repeated the same analysis as 
shown in Fig.~\ref{fig:allexponent} with these data, and Fig.~\ref{fig:comsolexponent} 
shows how the optimal exponent changes if data points at the smallest 
distances are progressively removed from the fitting. 
The optimal exponent obtained for the capacitance from the COMSOL simulation for 
the finite-size geometry deviates more from the ideal value of 0.5 as the cylinder-plane 
separation is increased ({\it i.e.} the regime in which edge effects are more pronounced). 
In contrast, the optimal exponent obtained experimentally in our electrostatic calibrations 
deviates more strongly from the ideal value as the separation is decreased. Therefore, border effects 
cannot explain, either quantitatively or qualitatively, the anomalous exponent 
observed in our measurements. 

\subsection{Local geometrical deformations}

Another possible reason for the deviation of the exponent with respect to the 
ideal value $2.5$ is the presence of geometrical deformations in the 
shape of the cylinder. As the experiment is performed with a very large cylinder 
($a$=12mm), the surface may present local deformations at the submillimeter scale, 
which could induce strong deviations in the exponent. A similar point has been 
raised for the case of a large sphere in front of a plane in \cite{DeccaComment}, 
where it was remarked that departures from the ideal spherical surface may 
noticeably affect the exponent of the electrostatic calibration.

\begin{figure}[t]
\includegraphics[width=0.45\columnwidth, clip=true]{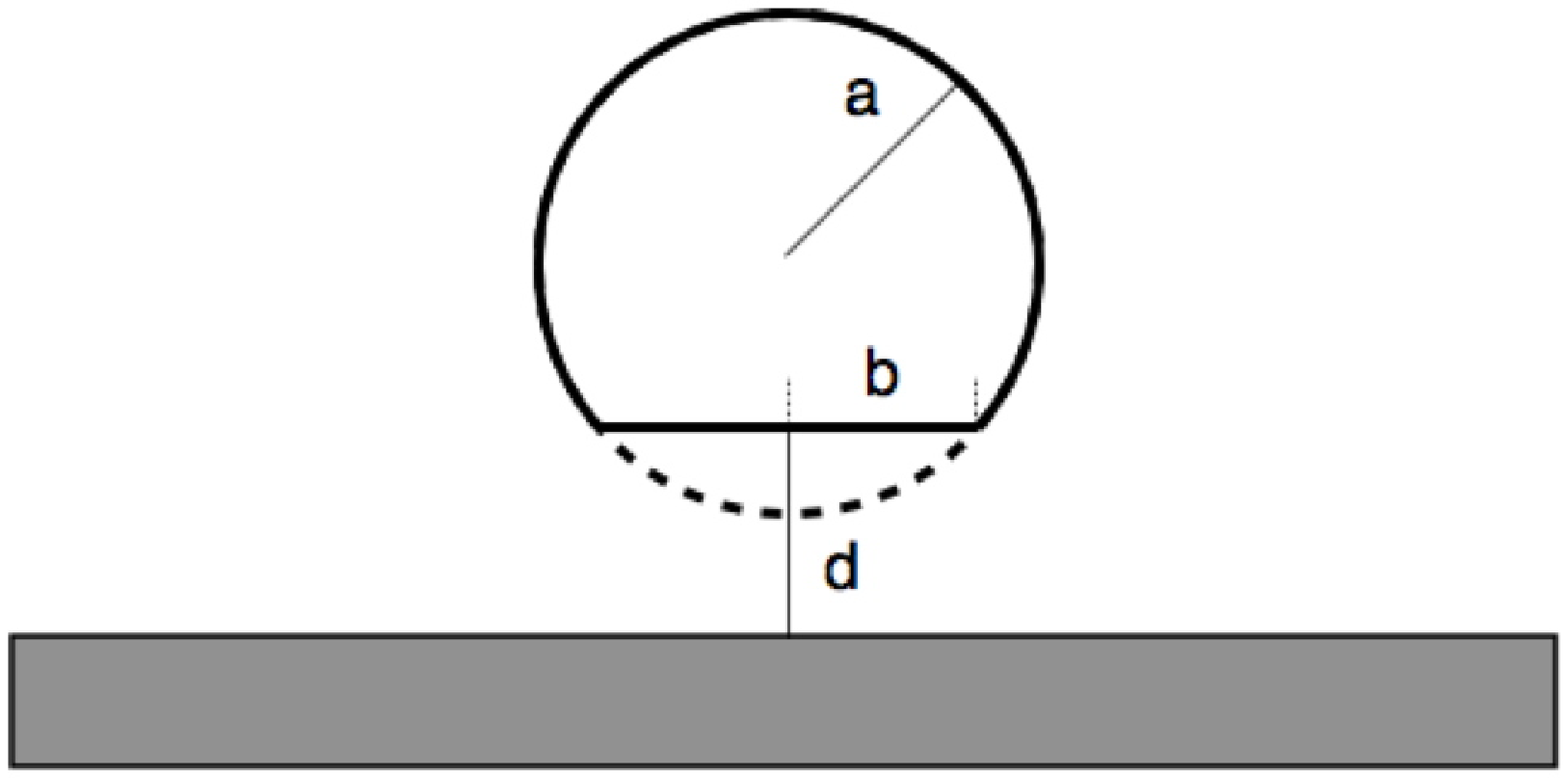}
\includegraphics[width=0.45\columnwidth, clip=true]{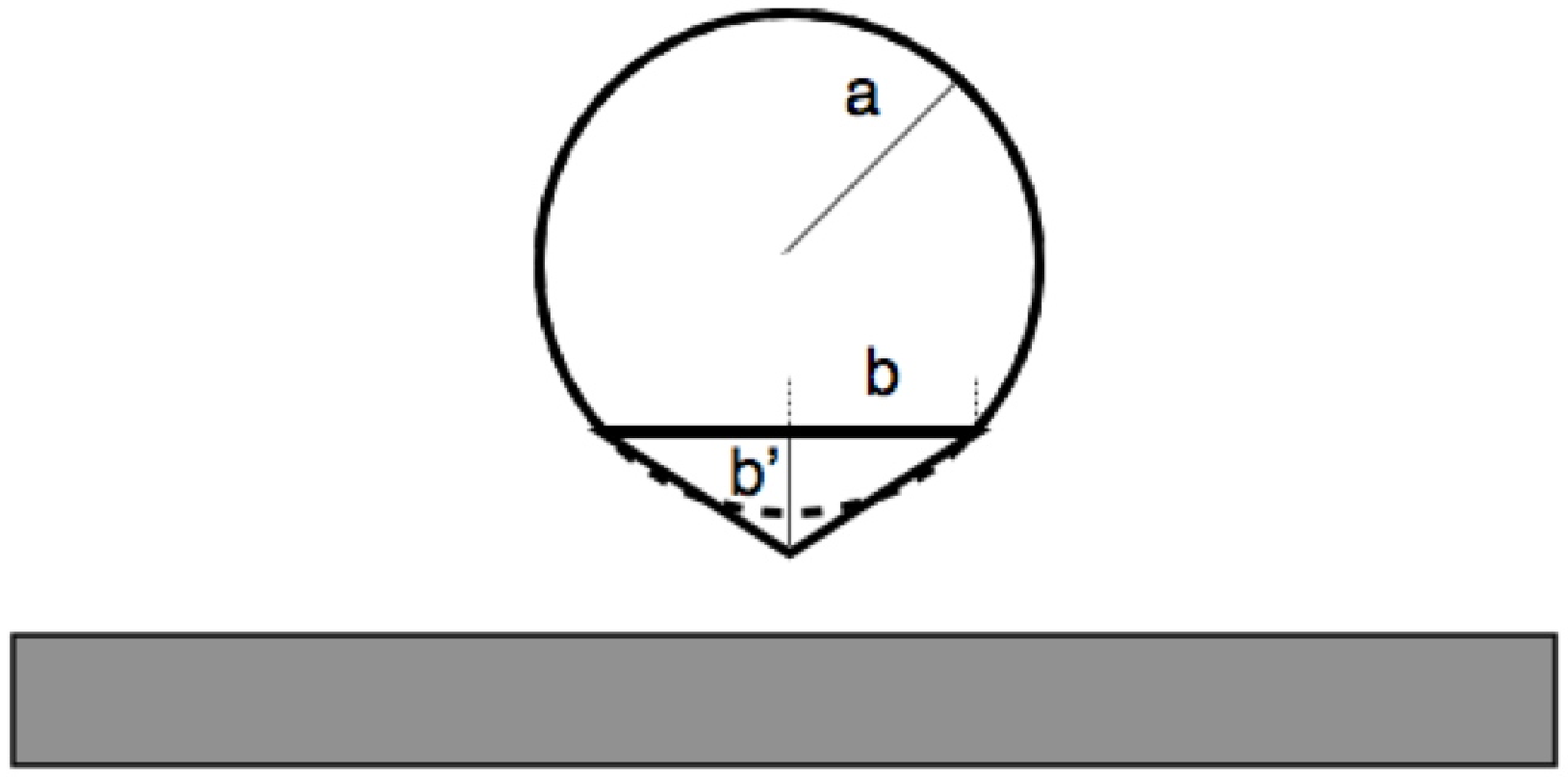}
\caption{Two examples of deformations of the cylindrical surfaces. On the 
left we show a cylindrical lens with a flat deformation, on the right, the 
cylinder has a tip (not to scale).} 
\label{tips}
\end{figure}

To illustrate this point, we have computed the electrostatic force between 
a deformed cylinder and a plane using the PFA. Let us first assume that, in the 
region of minimum distance between surfaces, the cylinder has a flat deformation 
of width $2{b}$ (see left plot in Fig. \ref{tips}). For simplicity we assume that 
the same deformation is all along the length $L$ of the cylinder. In this case 
it is possible to show that
\begin{equation}
\Delta\nu_\mathrm{el}^2= - \frac{\epsilon_0L_{\rm eff}(V-V_0)^2}{4\pi^2m_{\rm eff}}\frac{\partial^2}{\partial d^2} 
[f_{\rm inc}(d-\frac{{b}^2}{2a})+f_{\rm pp}(d)]\, , 
\end{equation}
where 
\begin{equation}
f_{\rm inc}(d)=\sqrt{\frac{2a}{d}} \arctan\sqrt{\frac{2ad}{{b}^2}}
\end{equation}
is the contribution of the (incomplete) cylinder, and  
$f_{\rm pp}(d)={b}/d$ is the contribution of the flat deformation. 
Let us assume that $\Delta\nu_\mathrm{el}^2= -A/d^B$. Depending on the 
relation between the size ${b}$ of the deformation and the range of 
distances $d$ of the calibration, we expect that the exponent $B$ will 
interpolate between the ideal value $2.5$ (cylinder-plane) at relatively 
large distances, and $3$ (parallel plates) at short distances (although the 
interpolation is not necessarily a monotonic function). When ${b}=10^2\mu$m,  
and the fit is performed for $d$ between 0.5 and 2 $\mu$m, the exponent 
becomes $B = 2.8$, bigger than that of the ideal cylinder-plane geometry. 
Then this kind of deformation does not help to explain the observed anomalous exponent.

On the other hand, if the cylinder has a deformation with the form of a tip (see right 
plot in Fig. \ref{tips}), the exponents are, in general, considerably smaller than 
the ideal one, and can explain at least part of the anomaly. 
Indeed, let us assume that the deformation consists of a triangular tip of width  
$2{b}$ and  height ${b'}$.  In this case, PFA gives 
\begin{equation}
\Delta\nu_\mathrm{el}^2= - \frac{\epsilon_0L_{\rm eff}(V-V_0)^2}
{4\pi^2m_{\rm eff}}\frac{\partial^2}{\partial d^2} [f_{\rm inc}(d+{b}')+f_{\rm tip}(d)]\, , 
\label{tip}
\end{equation}
where
\begin{equation}
f_{\rm tip}=\frac{{b}}{{b}'}\ln\left(1+\frac{{b}'}{d}\right)\, .
\end{equation}
If the height of the tip is much larger than $d$, the contribution of the incomplete cylinder is
almost irrelevant, since the cylinder is shifted upward and its electrostatic energy becomes
almost independent of $d$ in this regime. The main contribution comes from the tip, and 
the mild logarithmic dependence of the energy with the distance implies an exponent of around $B \approx 2$. 
This can be easily confirmed by performing fits of Eq. (\ref{tip}).
For example, for the same parameters as previously, with $b'=b$, we obtain an exponent $B = 2.0$. 
Sharper tips may produce even smaller exponents, although the PFA becomes unreliable for very thin tips. 

The main conclusion of the PFA estimations is that, as expected, deformations of the 
cylindrical surface may change appreciably the exponent $2.5$ of the electrostatic calibration.
Although the previous examples do not explain the full discrepancy between 
the ideal prediction and the experimental data, this is 
certainly a crucial point to be taken into account in future experiments.

\subsection{Additional electric forces}

\begin{figure}[b]
\includegraphics[width=0.5\columnwidth, clip=true]{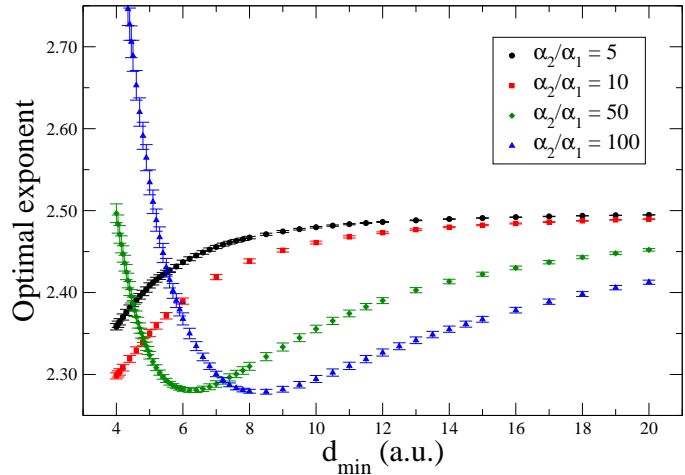}
\caption{(Color online) Optimal exponent of the hypothetical scenario data 
versus the distance of the closest data point used in the fitting. 
The data are constructed from Eq.~(\ref{eqn:nu_extrapower}) with 
$p$ = 5, $\alpha_1$ = 10${}^4$, and $\alpha_2/\alpha_1$ = 5, 
10, 50, 100 for the black circle points, red square points, green diamond points and 
blue triangle points, respectively.
}
\label{fig:extrafitexponent}
\end{figure}

\begin{figure}[ht]
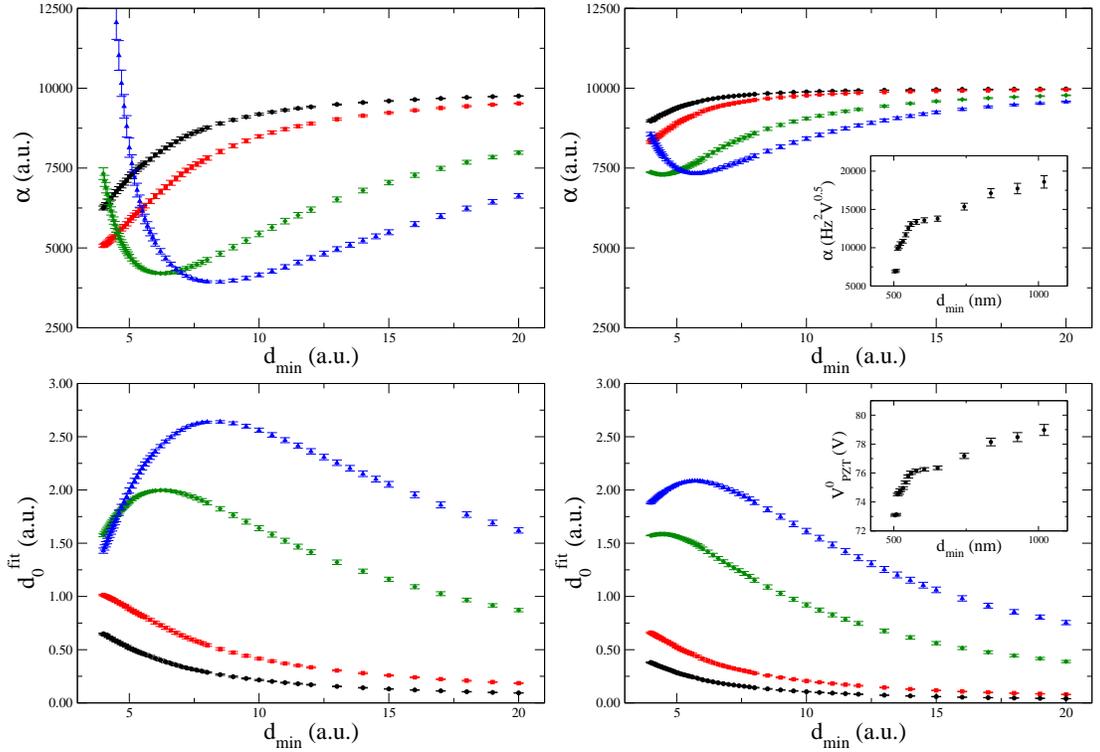

\includegraphics[width=0.4\columnwidth, clip=true]{cylindernew.fig17a.eps}
\includegraphics[width=0.4\columnwidth, clip=true]{cylindernew.fig17b.eps}
\includegraphics[width=0.4\columnwidth, clip=true]{cylindernew.fig17c.eps}
\includegraphics[width=0.4\columnwidth, clip=true]{cylindernew.fig17d.eps}
\caption{(Color online) Plots of the calibration factor $\alpha$ (top) and contact distance 
$d_0^\mathrm{fit}$ (bottom) of the hypothetical scenario data 
versus the distance of the closest data point used in the fitting. The plots on 
the left are the results of the best fit when the exponent $q$ is a free parameter, 
and the plots on the right are the results when $q$ is fixed at 2.5.
The data are constructed from Eq.~\ref{eqn:nu_extrapower} with 
$p$ = 5, $\alpha_1$ = 10000, and $\alpha_2/\alpha_1$ = 5, 10, 50, 100 for the black, 
red, green and blue curve, respectively.  The insets on the right
are the corresponding results using experimental data from our electrostatic 
calibration measurements. The relationship between the distance and the PZT voltage is 
$d - d_0 = \beta(V^0_\mathrm{PZT} - V_\mathrm{PZT})$ where $\beta$ is the actuation 
coefficient of the piezoelectric transducer. The same labeling for the points 
as in the previous figure is used.}
\label{fig:extrafit}
\end{figure}

In the residuals analysis of electrostatic calibrations described in the previous section, 
a residual force which seems to depend on the applied bias voltage was observed. The presence 
of such a force could lead to an anomalous exponent. 
Let us consider a hypothetical scenario in which the square of frequency shift $\Delta\nu_\mathrm{hs}^2$ 
for a cylinder-plane geometry has the following dependence on distance:
\begin{equation}
\Delta\nu_\mathrm{p}^2 =  - \left(\frac{\alpha_1}{d^{2.5}} + \frac{\alpha_2}{d^p}\right) (V-V_0)^2. 
\label{eqn:nu_extrapower}
\end{equation}in which $p > 2.5$. This means that besides the expected electrostatic 
interaction between the two surfaces, there is another electric force which follows a higher 
power law upon the cylinder-plane separation. Let us generate a set of pseudo-data following 
Eq.~(\ref{eqn:nu_extrapower}), and let us try to fit the curvature 
coefficient $K_\mathrm{p} = \alpha_1 d^{-2.5} + \alpha_2 d^{-p}$ with: 
\begin{equation}
K_\mathrm{p} = \alpha (d-d_0)^{-q}
\label{eqn:nu_extrafit}
\end{equation}
where $\alpha$, $d_0$ and $q$ are fitting 
parameters. In Figs.~\ref{fig:extrafitexponent} and ~\ref{fig:extrafit} we show 
the results of the fitting using different values of 
$\alpha_2/\alpha_1$ versus the distance of the closest point used in the fitting with $p$ = 5. 
Not surprisingly, the optimal value of $q$ from the fitting 
approaches $p$ at small distances, and approaches 2.5 at large distances. However, 
it is not monotonically decreasing from $p$ to 2.5 when the distance increases, but 
first goes below 2.5 and then slowly approaches 2.5. The $\alpha_2/\alpha_1$ = 5 and 10 
curves in Fig.~\ref{fig:extrafitexponent} resemble the results from our electrostatic 
calibrations data as shown in Fig.~\ref{fig:allexponent}. 
Therefore it is possible that an extra force with a power law dependence on 
distance steeper than the Coulombian one is responsible for the strong 
deviation of the optimal exponent from 2.5. The value of $\alpha$ 
from the fitting also suffers from similar issues. In the electrostatic calibration 
measurement, the exponent is fixed at 2.5 and the calibration factor $\alpha$ is 
used to calculate the effective mass $m_\mathrm{eff}$ of the resonator. 
However, when $\alpha_2 \neq 0 $ there is this extra force which is not 
included in the fitting formula, thus $\alpha$ obtained from the fitting is 
not equal to $\alpha_1$ as can be seen in the top plots of Fig.~\ref{fig:extrafit} ($\alpha_1$ is chosen to 
be 10${}^4$), and $m_\mathrm{eff}$ calculated from $\alpha$ would be incorrect. For example, 
the $\alpha_2/\alpha_1$ = 5 and 10 curves show that $\alpha$ obtained from the fitting 
is smaller than $\alpha_1$ = 10${}^4$. Since $m_\mathrm{eff}$ is inversely proportional 
to $\alpha$, a smaller $\alpha$ would result in a larger $m_\mathrm{eff}$. This is in  
agreement with our observation that the effective mass obtained from
electrostatic calibration measurements is larger than expected. The inset of the top right plot 
in Fig.~\ref{fig:extrafit} shows the calibration factor $\alpha$ versus the distance 
of the closest point used in the fitting with experimental data from our electrostatic 
calibration measurements, showing qualitative agreement with the model. 

The previous analysis indicates that with a carefully chosen combination of 
$p$ and  $\alpha_2/\alpha_1$, the existence of an extra electric force could well 
explain the problems we experienced in our electrostatic calibration measurements. 
Another very important result from this analysis is that 
unreliable values of $d_0$ could be obtained from the fitting if there exist 
forces other than the expected electrostatic force. As shown 
in the bottom plots of Fig.~\ref{fig:extrafit}, positive values of $d_0^\mathrm{fit}$ were 
obtained, both when the exponent $q$ is left as a free 
parameter or fixed at 2.5. Based on the way the data were constructed, the correct 
value for $d_0$ is 0 nm, and the difference between $d_0^\mathrm{fit}$ and its 
corresponding null value cannot be covered by the fitting uncertainty. 
This results in a further source of systematic error in the determination 
of the absolute distance, which adds up to other sources, including the 
one recently discussed in \cite{Zwol}. It should also be noted that 
there seems to be a strong correlation between the two fitting parameters 
$\alpha$ and $d_0^\mathrm{fit}$, which is indeed also present in our electrostatic 
calibration measurements as shown in the insets in the bottom right plot of  
Fig.~\ref{fig:extrafit}.

\subsection{Electrostatic patch effects}

As it is well known, all metallic surfaces in reality are not equipotential 
surfaces, showing instead voltage variations of order 10-100 mV over micrometer distances. 
These patch potentials are typically due to local changes in the work function 
associated to different crystallographic facets of the metal. 
Electrostatic patches are known to be an important systematic in several precision 
measurement experiments, including those aiming at detecting the Casimir force.
Apart from the static component, patch potentials can also fluctuate in time, a 
dynamical process that has not been studied in detail to date.  
It has recently been shown that by cooling a Au sample the 
electric-field noise above the metal is substantially reduced, 
a process possibly due to thermal activation barriers in the surface 
potential \cite{Chuang}.

Here we briefly describe the physics of electrostatic patches in the 
cylinder-plane geometry, and discuss whether they could be partly 
responsible for the anomalous exponent in our electrostatic calibrations. 
Our considerations follow closely the model and notation of \cite{Kimpatch2}. 
A related effect is the fluctuation-induced interaction between monopolar 
charge disorder within the dielectric slabs \cite{Podgornik}.
The electrostatic interaction energy between two parallel plates, whose surfaces
contain stochastic voltage variations $V_a(x,y)$ ($a$ denotes the upper or lower plate) is
\begin{equation}
U_{pp} = \frac{\epsilon_0}{2} \int_0^{\infty} dk \frac{k^2 e^{-k d}}{\sinh(k d)} S(k).
\end{equation}
This expression results from assuming zero-average patches, and an isotropic two-point 
correlation in the transverse plane-wave basis ${\bf k}$ given by 
$\langle V_{a,{\bf k}}, V_{b,{\bf k}'} \rangle= \delta_{a,b} C_{a, k} \delta^2({\bf k} - {\bf k}')$. 
Here $\langle \ldots \rangle$ denotes stochastic average, $k=|{\bf k}|$, and 
the power spectral density $S(k)$ is defined as 
$\int_0^{\infty} dk k S(k) \equiv (1/8 \pi) \int_0^{\infty} dk k (C_{1,k} + C_{2,k})$. 
The corresponding electrostatic force due to these potential patches is given 
by $F_{pp} = - \partial U_{pp} / \partial d$. To compute the patch effect 
on the force in the cylinder-plane configuration we make use of the PFA 
to treat the curvature of the cylinder, which is a good approximation 
in the limit $d/a \ll 1$. We do not impose any restriction on the 
typical size of the patches  ({\it i.e.} we leave $k d$ arbitrary). 
In the limit $d/a \ll 1$  the electrostatic force due to patches in the 
cylinder-plane configuration is 
\begin{equation}
F_{cp} = \frac{\pi \epsilon_0 L}{2 \sqrt{2}} 
a \left( \frac{d}{a} \right)^{1/2} \int_0^{\infty} dk \frac{k^3 e^{-2 k d}}{\sinh^2(k d)} S(k).
\end{equation}
Two simple limiting cases can be analyzed. 
In the large patch limit ($k d \ll 1$) the force is given by 
\begin{equation}
F_{cp} \approx \frac{\pi \epsilon_0 L}{2 \sqrt{2}} \frac{a^{1/2}}{d^{3/2}} V^2_{\rm rms} ,
\label{patches}
\end{equation}
with $V^2_{\rm rms} = \int_0^{\infty} dk k S(k)$. 
This expression is exactly equivalent to the r.h.s. of  Eq. (\ref{forceSmythe}) 
with $V^2$ replaced by $V^2_{\rm rms}$, as expected for large patches. 
In the small patch limit ($k d \gg 1$) the force is exponentially suppressed
because the patches are small and change sign rapidly, resulting in a vanishing 
net interaction between the plates. It should be noticed that the patch force 
depends on distance as an inverse power law (with exponent 1.5) only in the small-patch limit. 
The smaller the patches, the faster the decay (bigger exponent). 
More importantly for our purposes is the fact  that the electrostatic 
patch force (\ref{patches}) is {\it independent} of the applied voltage $V$ 
between the cylinder and the plane. Therefore, it cannot explain the anomalous 
exponent of the electrostatic calibration, that stems from the $V$-dependent 
contribution to the force. The patch force (\ref{patches}) is, instead, a background 
force that  could possibly show up in the analysis of the electrostatic residuals, 
that is, the force {\it after} the subtraction of the Coulomb-like $V$-dependent terms. 

It could be argued that the presence of strong electric fields in the 
gap between the plates may, in principle, redistribute the spatial configuration 
of the patches, and then the question is whether the force between the redistributed
patches depends on $V^2$ (note that if this were the case, there might be some hope 
that the anomalous exponent is partially due to patches). 
Equation (\ref{patches}) was obtained assuming that the two plates had only 
stochastic potentials fluctuating around zero. 
If an external fixed potential difference $V$ is applied between the plates, the 
linearity of Laplace's equation implies that the total force will be the sum 
of the usual $V^2$ term plus the $V$-independent term given in (\ref{patches}). 
In principle, for sufficiently large external fields in the gap, the power 
spectrum $S(k)$ could depend on the external voltage $V$, but it is unclear 
if, and how,  such an effect can account for the anomalous exponent in our 
electrostatic calibrations, at least in the ones performed by maintaining 
the electric-field approximately constant in the explored distance range.

\section{Minimizing potential in the parallel plates geometry}

\begin{figure}[t]
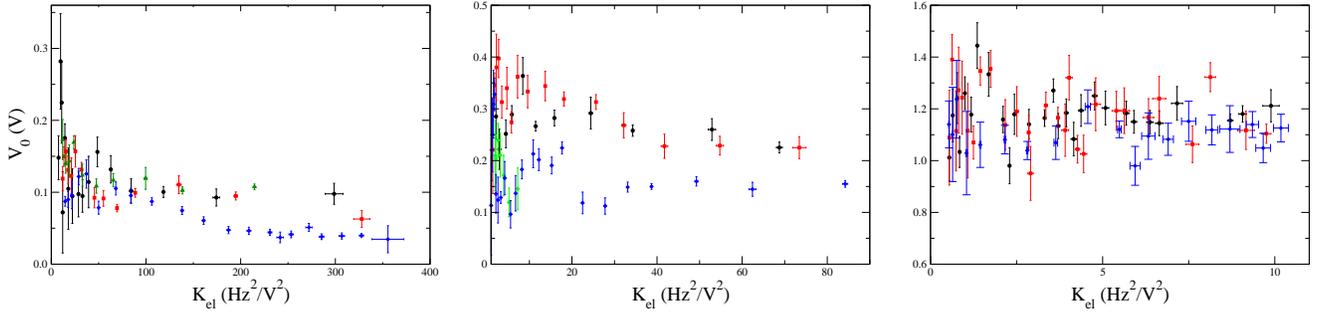

\includegraphics[width=0.32\columnwidth, clip=true]{cylindernew.fig18a.eps}
\includegraphics[width=0.32\columnwidth, clip=true]{cylindernew.fig18b.eps}
\includegraphics[width=0.32\columnwidth, clip=true]{cylindernew.fig18c.eps}
\caption{(Color online) Plots of the minimizing potential $V_0$ versus curvature coefficient 
$K_\mathrm{el}$ from various runs in plane-plane geometry with three different coatings of the 
mirror, Au (left), Ag (middle), and Al (right), all facing the Au-coated resonator.}
\label{fig:plates}
\end{figure}

We have also performed in the same experimental conditions measurements 
in the plane-plane configuration using flat coated mirrors facing the resonator.
Considering that the sphere-plane geometry has been the subject of former work 
\cite{PRARC,JPCS}, this allows us to complete the picture on the relationship between the 
distance-dependent minimizing potential and the specific geometrical configuration.
Three mirrors with different coatings (Au, Ag and Al) were used to also 
investigate the possible relationship between $V_0$ and the nature of the substrate. 
Although the parallelization is obtainable with a good level of approximation in one 
direction only, we can introduce an off-line parallelization correction in the data fitting. 
We analyze separately in the following results on the three substrates.
The data of $V_0$ from four runs were merged in the same plot for each of the three mirrors as shown in
Fig.~\ref{fig:plates} where they appear versus the curvature coefficient 
$K_\mathrm{el}$ instead of the distance. This is because 
the range of $K_\mathrm{el}$ achieved is not very large especially in the case 
of aluminum mirrors due to limited parallelization and smaller conductivity from 
oxide layers on the surface of the mirror, and consequently the usual fitting procedure to find 
the absolute distance would produce rather unstable and inaccurate results.
Since there is a one to one mapping between $K_\mathrm{el}$ and the distance, $K_\mathrm{el}$ 
is used here as a fair indicator of the distance for all runs. 

In the case of Au mirrors, $V_0$ fluctuates around 100 mV
without any noticeable trend. The fluctuation is bigger at large
distance partly because the frequency shifts tend to be small also 
for relatively larger bias voltage, this being reflected in the error 
bars in the fitting procedure. As the distance gets smaller, $V_0$ 
seems to converge to a constant value.  
The minimizing potential $V_0$ shows a similar 
behavior in all runs, and the values of $V_0$ are also comparable, 
with runs 1, 2, and 4 all around 0.1 V, and run 3 slightly lower at around 0.4 V. 
In Table III, the average value and the standard deviation of $V_0$ for each run 
is obtained using the six data points with largest $K_\mathrm{el}$ value from each run. 
We see that the standard deviation is reasonably small indicating that
$V_0$ is mostly constant.

\begin{table}[b]
\begin{center}
\begin{tabular}{|l|c|c|c||c|c|c|c|}
	\hline \hline
         & Run    &  $V_0$ (V)      &     $\sigma_{V_0}(V) $ & $\langle V_0 \rangle$ (V) & $W$ (eV) & $E_F$ (eV) & $E_F-W$ (eV) \\
	\hline
Gold     &  1     &  0.113      &     0.025  & 0.048 &5.1-5.47 & 5.53 & 0.06-0.43 \\
         &  2     &  0.089      &     0.017  &       &         &      &      \\
         &  3     &  0.041      &     0.006  &       &         &      &      \\
         &  4     &  0.113      &     0.008  &       &         &      &      \\
	\hline 
Silver   &  1     &  0.145      &     0.017  & 0.194 & 4.52-4.74 & 5.49 & 0.75-0.97 \\
         &  2     &  0.182      &     0.044  &       &           &      &      \\
         &  3     &  0.264      &     0.044  &       &           &      &      \\
         &  4     &  0.264      &     0.023  &       &           &      &      \\
        \hline 
Aluminum &   1    &  1.177      &     0.033  & 1.151 & 4.06-4.26 & 11.7 & 7.44-7.64 \\
         &   2    &  1.169      &     0.097  &       &           &      &      \\
         &   3    &  1.118      &     0.036  &       &           &      &      \\
\hline
\end{tabular}
\caption{Average measured value of $V_0$ for mirrors made of various 
substrates, weighted average value of $V_0$, and comparison with 
tabulated data for the work function $W$ and the Fermi energy $E_F$.}
\end{center}
\label{table:tableV0}
 \end{table}

Using Ag mirrors instead, a noticeable difference evident from the 
electrostatic measurement is that we achieve values of $K_\mathrm{el}$
smaller than in the case of Au mirrors. This is possibly due to the fact that 
Ag is easily oxidized once exposed in air prior to the insertion 
into the vacuum chamber. Of course, we cannot rule out the possibility that 
this could also be partly due to slightly different parallelization configurations. 
In terms of residual potential $V_0$, the Ag mirror shares the same behavior 
as gold mirror, as $V_0$ fluctuates randomly remaining constant especially 
at small separation gap. However, in this case the $V_0$ value is noticeably 
higher that that of the Au mirror. 

\begin{table}[b]
\begin{center} 
\begin{tabular}{|l|c|c|c|}
\hline\hline
         & Sphere-Plane   & Cylinder-Plane  & Parallel Planes  \\ 
\hline  
Casimir  &  $\frac{\pi^3}{360} \hbar c \frac{R}{d^3}$  
         &  $\frac{\pi^3}{384\sqrt{2}}\hbar c \frac{La^{1/2}}{d^{7/2}}$ 
         &  $\frac{\pi^2}{240} \hbar c \frac{S}{d^4}$    
\\ 
\hline 
Coulomb  & $\pi \epsilon_0 \frac{R}{d}V^2$
         & $\frac{\pi \epsilon_0}{2\sqrt{2}} \frac{La^{1/2}}{d^{3/2}}V^2$                
         & $\frac{\epsilon_0}{2} \frac{S}{d^2}V^2$      
\\ 
\hline 
$V_{\mathrm{Cas}}^{\mathrm eq}$(d) &
         $\left(\frac{\pi^2}{360}\right)^{1/2} \left(\frac{\hbar
         c}{\epsilon_0} \right)^{1/2} \frac{1}{d}$                
         &  $\left(\frac{\pi^2}{192} \right)^{1/2} \left(\frac{\hbar
         c}{\epsilon_0} \right)^{1/2} \frac{1}{d}$              
         & $\left(\frac{\pi^2}{120}\right)^{1/2} \left(\frac{\hbar%
         c}{\epsilon_0} \right)^{1/2} \frac{1}{d}$
\\ 
\hline 

$V_{\mathrm{Cas}}^{\mathrm{eq}}$ (1 $\mu$m) & 9.85 mV & 13.5 mV & 17.1 mV 
\\ 
\hline\hline
\end{tabular}
\end{center}
\caption{Summary of relevant formulas for the ideal Casimir force and the 
Coulomb force in the cases of the sphere-plane, cylinder-plane, and 
parallel plane geometries, with both forces in the first two geometries 
evaluated using PFA. In the third row the equivalent Casimir
voltage, {\it i.e.} the voltage which needs to be applied in order to 
simulate the Casimir force at a given distance $d$, is reported. 
In the last row the concrete value of the equivalent Casimir 
voltage is reported in the case of a typical gap distance of 1 $\mu$m.}
\end{table}

Finally, with an Al coating, the lowest value of $K_\mathrm{el}$ achieved 
is much smaller than that of the Au and Ag mirrors. 
This may be due to the easiness to form oxide layers, and it could 
also be related to the smaller conductivity with respect to Al and Au. 
In this case the residual potential $V_0$ is also constant at small separation 
gaps as it can be seen in Fig.~\ref{fig:plates}, and it is manifest that $V_0$ 
is significantly larger than that for Au and Ag mirrors. 
A more quantitative analysis is precluded by the strong dependence of the 
work function of Al on the exposure time in air \cite{Uda}.
  
A comparison among the various minimizing potentials $V_0$ and physical 
parameters of the substrate is shown in the second part of Table III. 
The average values of $V_0$ are obtained by averaging three runs from each 
mirror weighted by their variance. From the table it is clear 
that $V_0$ for mirrors coated with Au, Ag, and Al are significantly different. 
For a rough theoretical comparison, the work function and the Fermi
energy as well as their difference are also listed in the table. 
The work function of a metal is closely related to its Fermi energy, 
but due to the presence of defects and impurities on the  surface these 
two quantities do not coincide, with their difference largely due to 
the surface charge distribution and surface dipole distribution. 
From Table III we can see that Au has the smallest difference 
between its work function and Fermi energy while Al has the largest. 
Since this difference indicates the magnitude of the surface charge and dipole 
distribution which may be directly related to the residual potential $V_0$, it could 
be used to explain the different value of $V_0$ measured using mirror with different 
coatings. In fact the average $V_0$ values for Au, Ag, and Al mirrors is consistent 
with the order of this difference. Regarding the best fitting exponent for 
the scaling of the curvature coefficient with distance, we have 
obtained the expected one from Coulomb force, although it should 
be noticed that the range of distances was limited to a minimum value of about 3 $\mu$m.

\section{On the importance of the measurement and the modelization of the
minimizing potential in Casimir force measurements}

Our emphasis on measuring the minimizing potential at all the explored
distances for the sphere-plane configuration \cite{PRARC,JPCS}, and for
the cylinder-plane and parallel planes configurations described here,
is due to the fact that the understanding of the minimizing potential 
dependence on distance and time is crucial for the assessment of 
the accuracy of the Casimir force measurements at small (below $
\simeq 1 \mu$m) distances, and for determining the thermal 
contribution at large (in the 1-5 $\mu$m range) distances. 
In Table IV we  report the formulas for the ideal (perfect reflectors, 
zero temperature) Casimir force and for the Coulomb force in the three 
different geometries. Following \cite{Oncar}, let us define the equivalent 
Casimir voltage as the external bias voltage that can simulate the ideal Casimir
force. Due to the different scalings with distance of Casimir
and Coulomb forces, this equivalent Casimir voltage must be specified 
at each distance. It is, however, important to point out that the
difference between the various geometries is just a numerical factor 
which makes this equivalent voltage for the sphere-plane configuration 
about half the value of the parallel plane case (with the
cylinder-plane, as is customary, in between the two extreme cases 
even from this point of view). 
The formula for the equivalent Casimir voltage is 
\begin{equation}
V_{\mathrm{Cas}}^{\mathrm{eq}}(d) =
         \left(\frac{\pi^2}{\xi}\right)^{1/2} \left(\frac{\hbar
         c}{\epsilon_0} \right)^{1/2} \frac{1}{d},
\end{equation}
where $\xi$=360, 192, and 120 for the sphere-plane, cylinder-plane, and 
parallel plate configurations, respectively. 

In the numerical example presented in the last raw of Table IV, at a 
distance of 1 $\mu$m, which can be considered the borderline between 
the short-distance regime and the long-distance regime in which the 
thermal contribution starts to play a significant role, this equivalent 
Casimir voltage ranges between 10 and 17 mV depending on the geometry. 
At 3 $\mu$m, where the thermal contribution is expected to contribute 
as 10-20 $\%$ of the total force signal, with an absolute value still 
large enough to be detectable in various apparata, the equivalent Casimir 
voltage is three times smaller ({\it i.e.} between 3 and 6 mV). 
This voltage is of the same order of magnitude of the variation of 
the minimizing potential in a range of few micrometers. 
To take into account this contribution to properly 
subtract it from the data, it is therefore necessary to model 
the minimizing potential at the few percent level accuracy. 
Unfortunately, such stringent theoretical characterization of the 
minimizing potential is not yet available. 
  
A further degree of uncertainty is also related to the fact that, as 
pointed out in \cite{JPCS}, the minimizing potential may depend on time, 
a fact that, for instance, could be attributed to temperature drifts. 
It has been experimentally shown in \cite{Remmert} using a heated 
atomic force microscope tip that the contact potential depends  
on temperature, with a slope estimated to be of the order of 4 mV/$^0$C.
To perform high-precision tests of the Casimir force, one 
therefore needs a stringent temperature stability of the apparatus 
during the entire measurement run. In \cite{Pollack}, the observation 
of fast changes in the contact potential have been conjectured as due 
to the effect of background cosmic rays impinging on the apparatus. 
For a release of about 10$^{-11}$ C/cm$^2$ through ionization by cosmic 
rays at sea level, and considering the small values of the capacitances 
(order of hundreds pF), sudden changes of order 0.1-1 mV 
could be expected (see \cite{Mitro} for a related discussion). 
A careful control of the minimizing potential will then require 
also surrounding the apparatus with lead shields to reduce the 
radiation background or particle detectors to veto the apparatus 
during large ionization events, especially for apparata using microspheres.

\section{Conclusion}

In this paper we have summarized the main outcomes from our effort 
to measure the Casimir force in the cylinder-plane configuration. 
The presence of uncontrollable frequency shifts of electric origin at 
the smallest explored gaps, evidenced both from the analysis of the 
residuals of the electrostatic calibrations and a fast-approach
technique, prevent us from identifying a Casimir-like contribution at small
distance. At large distances thermal drifts are large, if 
compared to the expected Casimir force, and a careful control 
of the dependence of the minimizing potential on distance is 
also required to extract meaningful information about the 
Casimir force and its thermal corrections. 
While unsuccessful, our search for the Casimir force in this geometry, 
apart from the development of some data taking and analysis 
techniques applicable elsewhere, has evidenced a number 
of features which may be of more general interest, as we try to 
summarize in the following. 

First, we have observed anomalous behavior for the best fitting
exponent with which the electrostatic coefficient is scaling as a function 
of the cylinder-plane separation. The exponent is significantly smaller 
than the Coulombian one at small distances, while it retains its 
expected value at the largest explored gaps.
In the case of the sphere-plane measurements, the exponent was 
slightly smaller than the expected value but it retained its value
in the entire explored range of distances, this last being smaller 
than in the cylinder-plane case due to the smaller electrostatic 
signal available in the sphere-plane configuration \cite{PRARC}. 

Second, we have observed a dependence of the minimizing
potential on the cylinder-plane distance, similarly to 
the sphere-plane case. Although its dependence is milder at small 
distances, it still retains a strong dependence on distance at 
larger gaps, and this requires a careful modeling 
to subtract its contribution when studying forces in the 1-5
$\mu$m range of interest to discriminate among the various models 
proposed to incorporate the thermal contribution.

Finally, we have also explored the case of flat surfaces 
with a rough parallelization and we have found 
that in this case no anomalous behavior is observed 
for both the scaling exponent and the minimizing potential, even 
using different substrates for the surfaces. The range of explored 
distances is definitely limited by the approximate two-dimensional 
parallelism achievable with our setup, and it is therefore unclear if anomalous 
scaling could be instead observed as in the case of the smaller 
explored gaps in the sphere-plane and cylinder-plane configurations.
Both apparata built to study the Casimir forces in a parallel plane 
configuration have not observed dependence of the minimizing potential
on distance \cite{Bressi,Antonini2}.

Our findings should be then related to recent outcomes from various 
experiments confirming the presence of nontrivial, formerly unidentified 
systematic effects in the electrostatic calibrations \cite{PRARC,JPCS,Iannuzzi,Kimpatch1}.  
Several recent experiments are also showing that the observation 
of Casimir or Casimir-Polder forces is less trivial than previously 
stated, for instance, with regards to the dependence on the optical 
properties of the substrates \cite{Sveto1} and the presence of 
dielectric layers on the substrates \cite{Pala1}. 
Theoretical arguments have been recently provided for the nontrivial 
interplay between thermal fluctuations and geometry \cite{Giesrecent}, 
thermal, conductivity, and roughness corrections \cite{Paulo,Paulo1} and the 
role of the statistical  properties of the conducting surfaces
\cite{Podgornik}. Deviations from the pure Coulombian contribution
and from the hypothesis of a constant minimizing potential have also been  
observed in atomic force microscopy for sharp tips, for instance, due to 
capillary forces \cite{Jang,Huber}.

\acknowledgments
The work of DARD was funded by DARPA/MTO's Casimir Effect Enhancement program 
under DOE/NNSA Contract DE-AC52-06NA25396, and the work of FCL and FDM was 
supported by UBA, CONICET, and ANPCyT (Argentina).

\end{document}